\documentclass[useAMS,usenatbib,epsfig, a4paper]{mn2e}
\usepackage{epsfig}
\usepackage{times}
\usepackage{amssymb}

\newcommand{\ba}{\begin{eqnarray}}
\newcommand{\ea}{\end{eqnarray}}

\def\gtorder{\mathrel{\raise.3ex\hbox{$>$}\mkern-14mu
             \lower0.6ex\hbox{$\sim$}}}
\def\ltorder{\mathrel{\raise.3ex\hbox{$<$}\mkern-14mu
	                  \lower0.6ex\hbox{$\sim$}}}

\title[Map-making in small field modulated CMB polarisation experiments]{Map-making in small field modulated CMB polarisation experiments: approximating the maximum-likelihood method}
\author[D.~Sutton et al.]{D.~Sutton $^1$, B.R.~Johnson $^1$, M.L.~Brown $^2$,  P.~Cabella $^{1,3}$, P.G.~Ferreira $^1$, and K.M.~Smith $^6$\\ 
$^1$ Oxford Astrophysics, University of Oxford, Denys Wilkinson Building, 1 Keble 
Road, Oxford OX1 3RH, United Kingdom \\
$^2$ Cavendish Laboratory, University of Cambridge, J J Thomson Avenue, Cambridge, CB3 0HE, United Kingdom  \\
$^3$ Universita di Roma `Tor Vergata', Dipartimento di Fisica, via della
ricerca scientifica 1,  00133 Rome, Italy\\
$^4$ Institute of Astronomy,
University of Cambridge, 
Madingley Road,
Cambridge. CB3 0HA,
United Kingdom
}
\date{\today}
\pubyear{2008}

\begin{document}

\maketitle

\begin{abstract}
Map-making presents a significant computational challenge to the next generation of kilopixel
CMB polarisation experiments. Years worth of time ordered data (TOD) from thousands
of detectors will need to be compressed into maps of the T, Q and U Stokes
parameters. Fundamental to the science goal of these experiments,  the observation of B-modes, is the ability to control noise and systematics.  In this paper, we consider an alternative to the maximum-likelihood method, called \emph{destriping}, where the noise is modelled as a set of discrete offset functions and then subtracted from the time-stream.  We compare our destriping code (Descart: the DEStriping CARTographer) to a full maximum-likelihood map-maker, applying them to 200 Monte-Carlo simulations of time-ordered data from a ground based, partial-sky polarisation modulation experiment.  In these simulations, the noise is dominated by either detector or atmospheric $1/f$ noise.  Using prior information of the power spectrum of this noise, we produce destriped maps of T, Q and U which are negligibly different from optimal. The method does not filter the signal or bias the E or B-mode power spectra.  Depending on the length of the destriping baseline, the method delivers between 5 and 22 times improvement in computation time over the maximum-likelihood algorithm.   We find that, for the specific case of single detector maps, it is essential to destripe the atmospheric $1/f$ in order to detect B-modes, even though the Q and U signals are modulated by a half-wave plate spinning at $5$-Hz.

\end{abstract}


\section{Introduction}
\label{sec:intro}
Measurements of the temperature and polarisation anisotropy of
the cosmic microwave background (CMB) have been used to constrain
a number of cosmological parameters to high precision (\citealt{smoot:1992}, \citealt{hanany:2000}, \citealt{masi:2006}, \citealt{hinshaw:2008}).

There
are, however, still open questions that haven't been answered with the
new results. In particular there is still no evidence for or against the
existence of primordial gravity waves left over from a period of inflation
at very early times. A bath of gravitational radiation should leave its
imprint on the polarisation of the CMB in the form of a unique B-mode pattern. 
Such a signal will be faint compared to the polarisation that arises from density perturbations. 

If we are to make a convincing measurement of B-modes 
(or rule out their existence), we need to construct an experiment
with far greater sensitivity than the current state of the art. Given the physical limitations on the
properties of individual detectors, the only realistic way of doing this
is by observing the sky with multiple detectors for long periods of time.
Most experiments that are being proposed have these characteristics.

A notable example is the C$_{\ell}$OVER experiment \citep{north:2008}. This experiment will have 576 detectors and will focus on small fractions of the sky for periods of up to two years. With
current planned technology, it is hoped that the noise per map pixel
in the experiment will be below 0.8-$\mu$K.  To achieve this, it is essential to supplement the increased number of observations that multiple detector pixels bring by removing correlated noise in the map-making stage.

Solutions to the map-making problem aim to compress terabytes of time ordered data (TOD), acquired over months or years of scanning the sky, into a pixelised sky map, with noise uncorrelated between sky pixels.   In each time ordered datum, the signal is typically swamped by noise, necessitating the process of binning and averaging at the heart of every map-making algorithm.  Unfortunately, simple averaging is highly sub-optimal for this task because the noise in the TOD is correlated due to $1/f$ noise.  One could attempt to high-pass filter the resulting low frequency $1/f$ noise drifts, thereby leaving random uncorrelated or \emph{white} noise in the TOD.  Such a method is not lossless, as it filters the signal as well.  Whilst the effects of filtering can be mitigated in multipole space, by simulating its effects using Monte-Carlo simulations \citep{hivon:2002}, the variance of the recovered power spectra are larger.

A number of previous experiments, including COBE, MAXIMA and BOOMERANG, have opted to use the optimal maximum likelihood map-making algorithm (eg: \citealt{smoot:1992}, \citealt{hanany:2000}, \citealt{masi:2006}).  This algorithm has been well discussed in the literature (eg: \cite{tegmark:1997a}, \cite{natoli:2001}, \cite{stompor:2002}, \cite{de-gasperis:2005}) and is renowned for being slow and memory intensive.  In addition, the algorithm requires accurate knowledge of the noise statistics of the TOD.  Use of this algorithm is impossible for the next generation of high resolution, kilo-pixel, polarisation experiments, where we will require Monte-Carlo noise maps to reconstruct the noise covariance required for estimating B-modes on the cut sky \citep{ksmith:2007}.

Destriping is a promising alternative to the maximum likelihood approach (\cite{keihanen:2004}, \cite{keihanen:2005}).  By modelling the correlated noise in the TOD as a series of offset functions mapped onto the time-series, a nearly optimal map can be returned very quickly by solving for the offset function amplitudes. This system is much smaller than the full maximum likelihood system and can therefore be solved at a fraction of the computational cost.  Destriping has been well discussed in the literature for the Planck experiment and favourably compared, in simulations, to the optimal solution: destriping can produce near-optimal maps in orders of magnitude less time (\citealt{poutanen:2006}, \citealt{ashdown:2007a}, \citealt{ashdown:2007b}).  A variant of destriping has been successfully used in the Archeops balloon experiment analysis pipeline \citep{macias-perez:2007}.  However, these experiments are very different to upcoming ground-based B-mode experiments in terms of scanning strategy, presence of atmospheric noise and in the number of detectors.  

Planck and Archeops use a \emph{circular} scanning strategy: scanning the CMB in overlapping great circles before re-pointing to scan a new great circle.  Scanning strategies are very different for ground based experiments, as a matter of necessity, because the scanning must be in azimuth only (constant elevation) in order to avoid extremely high atmospheric noise contamination from changing elevation.   This causes a large difference in the level of cross-linking, which has a strong effect on the performance of map-making.

The biggest difference between ground based B-mode experiments and Planck or Archeops is the presence of atmosphere (Archeops flew so high it was essentially \emph{above} the atmosphere) .  Atmospheric $1/f$ noise dominates the correlated noise in the TOD, even for the current generation of ground based experiments. It has a higher knee frequency and higher spectral index than the detector noise.  For example, raw (undifferenced) TOD from the QUaD experiment \citep{hinderks:2008} appears to exhibit $1/f^{2}$ atmospheric noise rather than simple $1/f$.  Furthermore, the long term noise drifts will be very correlated between detector pixels \citep{bussman:2005}.

In this paper, we concentrate on experimental designs that modulate the faint polarisation signal to frequencies higher than the detector $1/f$  knee frequency $f_{knee}$, using a rotating half-wave plate, so that Q and U will be sampled in the white noise regime.  Upcoming experiments with this design include C$_{l}$OVER \citep{north:2008}, EBEX \citep{oxley:2005} and SPIDER \citep{mactavish:2007}.  These modulation experiments aim to mitigate possible detector $1/f$ using hardware.  However, sophisticated map-makers will still be required to produce optimal Q and U maps.  If we are to measure B-modes, we must treat the systematics in the data properly.  One of these systematics will be instrumental polarisation, which causes $T \rightarrow P$ leakage in the data \citep{johnson:2007}.  If the leakage is significant, it will be vital to remove the leaked temperature signal, a process that will require a high-resolution, near-optimal map of the temperature anisotropies as observed by the instrument.  Such leakage may also mix atmospheric $1/f$ noise into the Q and U data, for which a sophisticated map-maker like destriping will be needed to remove leaked $1/f$ in demodulated Q and U time-streams.  The atmospheric noise itself may also contain a polarised component that would likewise have to be removed in the map-making \citep{hanany:2003}.

This paper is the first in a series evaluating destriping as applied to ground-based, partial-sky, experiments.  We examine the speed and accuracy of destriping for modulated experiments using simulations, dominated either by detector or atmospheric correlated noise, of a single detector pixel.  This is akin to assuming that the noise is uncorrelated between detectors. We note that this is a poor assumption for the atmospheric $1/f$ component: we will generalise from this assumption in the next paper in the series (see also Appendix \ref{multi-detector appendix}).

In Section \ref{algorithms}, we put our discussion on firm footing by deriving the optimal and destriping solutions to the map-making problems and discussing details of the implementation of the algorithms.  In Section \ref{simulations}, we describe the simulation of the TOD and the scanning strategies.  In Section \ref{results}, we first apply the algorithms to a preliminary signal-only simulation to test for bias and then proceed to apply the methods to the full Monte-Carlo simulations of TOD containing both signal and noise, comparing the performance of the algorithms directly with emphasis on speed and accuracy.

\section{The Algorithms}
\label{algorithms}

Time ordered data, denoted by a time vector $y_{t}$, is formed from the sum of two time-space  vectors, the sky signal $S_{t}$ and total noise per observation $n_{t}$.  The signal vector is created by the scanning of a telescope across the sky, which we consider to be innately pixelised.  With this assumption, the signal stream becomes the result of the action of a  ($N_{time} \times N_{pixel}$) projection operator, $P_{tp}$, on the sky map $x_{p}$.

This can be described in tensor notation by

\begin{equation}
y_{t} = P_{tp} x_{p} + n_{t}
\label{tod hard}.
\end{equation}

The reverse operation of $P_{tp}$ is the binning operator $\mathbf{P^{T}}= P_{pt}$ (where the superscript T denotes the transpose), which sums the TOD into a map.  The two operators acting together, $P_{pt} P_{tp}$,  sum the \emph{number of observations} per map pixel.

In the case of no noise, the map could be exactly returned by averaging: binning the TOD into a map and dividing each pixel by how many times it was observed

\begin{equation}
\vec{x_{p}}= (P_{pt'}P_{t'p})^{-1} P_{pt} \vec{y_{t}}
\label{naive estimator},
\end{equation}
which also returns the minimum-variance map in the case that the noise $n_{t}$ is white.  This defines the simplest  ``naive" map-making algorithm.

With polarisation measurements, we can measure 3 CMB skies: the
unpolarised (or temperature) sky $T$ and two skies corresponding to the $Q$ and $U$
polarisation parameters.  In this case, the sky map vector becomes

\begin{equation}
\vec{x} = 
\left( \begin{array}{c}
\vec{T} \\
\vec{Q} \\
\vec{U}
\end{array} \right),
\end{equation}
a vector of length ($3 \times N_{pixel}$), which has covariance matrix

\begin{equation}
\mathbf{C} = \langle \vec{x^{T}} \vec{x} \rangle
= \left( \begin{array}{ccc}
\langle \mathbf{TT} \rangle \langle \mathbf{TQ} \rangle \langle
\mathbf{TU} \rangle \\
\langle \mathbf{QT} \rangle \langle \mathbf{QQ} \rangle \langle
\mathbf{QU} \rangle \\
\langle \mathbf{UT} \rangle \langle \mathbf{UQ} \rangle \langle
\mathbf{UU} \rangle
\end{array} \right)
\label{polarized cov matrix}.
\end{equation}
with dimensions ($3 N_{pixel} \times 3 N_{pixel}$).

The projection operator $\mathbf{P}$ describes the modulation of the Q and U signals into the TOD, expanding into T,Q and U polarisation pointing matrices

\begin{equation}
\mathbf{P} = 
\left(\begin{array}{c}
\mathbf{P_{T}} \\
\mathbf{P_{Q}}\\
\mathbf{P_{U}}
\end{array} \right)
\label{P into polarisation matrices}.
\end{equation}

The half-wave plate experimental design modulates the Q and U signals into the time stream at a frequency $\omega = 4f$, where the modulation frequency $f$ is the frequency of half-wave plate rotation.  The modulation frequency is chosen to be larger than the knee frequency of the correlated $1/f$ noise, such that the low intensity Q and U signals are sampled in the white noise regime, at a frequency $\omega > 4 f_{knee}$.

The signal part of the modulated time stream is then
\begin{equation}
S_{t} = \frac{1}{2} P_{tp} \big(T_{p} + Q_{p} \cos{4\beta_{t}} + U_{p} \sin{4\beta_{t}} \big)
\label{mod_timestream}.
\end{equation}
where $\beta$ is the orientation angle of the half-wave plate with respect to the sky and $P_{tp}$ is the simple pointing matrix from (\ref{tod hard}) \citep{johnson:2007}.

This gives us the polarisation pointing matrices from (\ref{P into polarisation matrices})

\begin{eqnarray}
\mathbf{P}_{T} & = & \frac{1}{2} P_{tp} \nonumber  \\
\mathbf{P}_{Q} & = & P_{tp} \big( \frac{\cos{4\beta}}{2} \big)\nonumber \\
\mathbf{P}_{U} & = & P_{tp} \big( \frac{\sin{4\beta}}{2} \big)
\label{pol_point_hwp}.
\end{eqnarray}

\subsection{The Maximum-Likelihood Solution to the Map-Making Problem} 
\label{mle}

To estimate the sky map $x_{p}$, we begin with Bayes' theorem

\begin{equation}
P(\theta | D I) \propto P(D | \theta I) \times P(\theta | I),
\end{equation}
where $\theta$ is our list of parameters, $D$ is the data and $I$ is all of the assumed background information of the experiment.  In the map-making problem, the data is the TOD vector $y_{t}$ and the parameters are the sky map $x_{p}$ and the noise $C_{N,tt'}= \langle n(t) n(t') \rangle$.
Bayes theorem now becomes

\begin{equation}
P(x_{p} C_{N,tt'} | y_{t} I) \propto P(y_{t} | x_{p} C_{N,tt'} I) \times P(x_{t} | I) P(C_{N,tt'} | I).
\end{equation}

We assume that the noise is a stationary Gaussian realisation of $N(f)$, an underlying noise power spectrum, such that $\langle n_{t} \rangle = 0$, $C_{N_{tt'}} = C_{N}(t-t')$. Furthermore, we assume complete \emph{a priori} knowledge of the noise power spectrum $N(f)$.  This is like assuming a delta function prior on the noise, which effectively becomes part of the background information.  With a flat prior on $x_{p}$, the posterior is simply proportional to the likelihood

\begin{equation}
P(\vec{x} | \vec{y} I) \propto exp \big(-\frac{\chi^2}{2}\big),
\end{equation}
with
\begin{equation}
\chi^2 = (\vec{y} - \mathbf{P} \vec{x})^{T} \mathbf{C_{N}^{-1}} (\vec{y} - \mathbf{P} \vec{x})
\end{equation}
where we have switched from tensor to matrix notation.

Solving for $\vec{x}$ by minimising $\chi^{2}$, we
find

\begin{equation}
\vec{x} = (\mathbf{P^{T} C_{N}^{-1} P})^{-1} \bf{P^{T} C_{N}^{-1}} \vec{y}
\label{sky map},
\end{equation}
which has pixel noise covariance matrix (obtained from the second derivative $\partial^{2} \chi^{2}/\partial \vec{x}^{2}$)
\begin{equation}
\bf{C} =  (\mathbf{P^{T} C_{N}^{-1} P})^{-1}
\label{covariance}.
\end{equation} 

The matrix ($\bf{P}^{T}\bf{C_{N}}^{-1} \bf{P}$), sometimes called the \emph{weight matrix}, is large and sparse.  Traditional construction and inversion is prohibitive, with the memory requirement of storage scaling as $\mathcal{O}(N_{pixel}^{2})$ and explicit inversion methods scaling as $\mathcal{O}(N_{pixel}^{3})$.  Fortunately, there are a number of computational tricks at our disposal.

The first trick is used to make the computation of $\mathbf{C_{N}^{-1}}$ feasible.  $\mathbf{C_{N}}$ is an $(N_{t} \times N_{t})$ matrix whose explicit construction and inversion is impossible.  However, if we assume that the $1/f$ noise in the TOD is \emph{stationary},  the ensemble average matrix $\mathbf{C_{N}}$ can be well approximated as \emph{circulant} (where each row is exactly the same as the row above, all shifted one column to the right).  Systems involving circulant matrices can be inverted by deconvolution.  If $\mathbf{A}x = b$, where $\mathbf{A}$ is circulant, then $x= \mathcal{F}^{-1}[\mathcal{F}[b]/\mathcal{F}[\mathcal{A}_{1n}]]$, where $\mathcal{F}$ and $\mathcal{F}^{-1}$ denote Fourier  and inverse Fourier transformation respectively and $\mathbf{A}_{1n}$ is a vector formed of the first row of matrix $\mathbf{A}$.

This assumption is not valid for real data, as the TOD would have to wrap around to correlate the ends of the time-stream.  For noise simulated by Fourier transforms however, the circulant approximation is exact.  We note that for real data, the noise covariance is in fact a symmetric Toeplitz matrix, whose inverse can be calculated expensively in $\mathcal{O} (N_{t}^{2})$ operations using Levinson's method (see eg: \citealt{press:2002}). The standard approach is to use the ``MADCAP approximation" when inverting this matrix: the matrix is inverted as if it were circulant and then elements far from the diagonal are set to zero to enforce a Toeplitz structure, $C_{N}^{-1}(t-t') = 0$ for $|t-t'| > N_{t}/2$ \citep{borrill:1999}. This approximate inverse covariance can then be convolved with the TOD using FFTs.

For the noise covariance of our simulations, we have that

\begin{equation}
\mathcal{F}[\mathbf{C_{N}}_{1t}] = P(f)
\label{noise and power spectrum},
\end{equation}
where $P(f)$ is the ensemble average noise power spectrum and $\mathbf{C_{N}}_{1t}$ is the first row of $\mathbf{C_{N}}$.

The second trick speeds up the inversion of the weight matrix itself.  Rather than inverting the matrix explicitly, we solve the system in (\ref{sky map}) \emph{iteratively}, using an algorithm from the conjugate gradients family.  A number of maximum likelihood implementations have used a preconditioned conjugate gradients (pcg) algorithm for this task (eg: ROMA \citep{natoli:2001}, MapCUMBA \citep{dore:2001} and SANEPIC \citep{patanchon:2007}).  However, we implement a more robust algorithm from the conjugate gradients family: MINRES, an algorithm capable of solving any invertible symmetric matrix, regardless of positive definiteness \citep{barret:2006}.  By definition, the matrix is symmetric and should be invertible if there is a sufficient dispersion of polarisation measurement angles for each pixel.  The weight matrix should, analytically speaking, be positive definite.  We implement MINRES to protect against numerical departures from positive definiteness that can propagate from inaccuracies in noise estimation \citep{natoli:2001}.

The speed of iterative inversion depends on the condition number of the matrix (equivalent to the absolute ratio of the largest and smallest eigenvalues of the matrix).  A poorly conditioned system is not well posed for iterative inversion, so we insert a preconditioning step into the algorithm.  The system $\mathbf{A} x = b$ can be preconditioned using \emph{preconditioner matrix} $K$ as follows

\begin{equation}
K_{1}^{-1} A K_{2}^{-1} (K_{2} \vec{x}) = K_{1}^{-1} \vec{b}
\label{preconditioning}
\end{equation}
where the left and right hand preconditioners $K_{1}$ and $K_{2}$ are defined by $K_{1}K_{2}=K$.  The preconditioner should be chosen such that $K^{-1} \approx \mathbf{A}^{-1}$, but it should also be quick to compute.  In practice, choosing $K$ is a trade off between improvements in convergence and added computation time.  The size of the system prohibits the use of a sophisticated preconditioner, such as an incomplete Cholesky factorisation, so we use the simple but effective point Jacobi preconditoner.  This is simply the diagonal of the weight matrix, which we approximate as

\begin{equation}
K_{pp'} = \sum_{t} \sum_{t' = t-\lambda_{C}}^{t+\lambda_{C}} (P_{pt} C_{N, tt'}^{-1} P_{t'p'}) \delta_{pp'},
\label{preconditioner explicit},
\end{equation}
for a correlation length $\lambda_{C}$ ($\delta_{pp'}$ is the Kronecker delta).

Iterative inversion has a further advantage.  The explicit calculation of the weight matrix is not required, we only need to reproduce the action of its matrix multiplication on the sky map $\vec{x}$.  This is easily done with the following algorithm (in which the weight matrix is factorised into its component operations)

\begin{equation}
(\mathbf{P^{T} C_{N}^{-1} P}) \vec{x} = P^{T}_{tp} \mathcal{F}^{-1}\Bigg[\frac{\mathcal{F}\big[P_{tp}\vec{x}\big]}{\mathcal{F}\big[\mathbf{C_{N}}_{1t}\big]}\Bigg]
\label{prewhiting matrix mult},
\end{equation}
a process of projection, deconvolution (using fast Fourier transforms) and binning.  This final trick has decreased the memory requirement for the matrix from $\mathcal{O}(N_{pixel}^{2})$ to $\mathcal{O}(N_{t})$.

It should be noted that the computation time for the optimal solution is still unfeasible for future data sets, despite the tricks implemented into the algorithm.  The operational scaling of the algorithm is $\mathcal{O}(N_{it}\times N_{t}(1+ \log_{2}|N_{t}|))$, where $N_{it}$ is the number of iterations.  The binning and projection operators ($P^{T}_{tp}$ and $P_{tp}$) scale as $\mathcal{O}(N_{t})$ so that each iteration is dominated by the FFT and inverse FFT (both of which scale as $\mathcal{O}(N_{t}\log_{2}|N_{t}|)$).

\subsection{An alternative: the destriping method}
\label{destriping}

\begin{figure}
\begin{center}
\includegraphics[angle=0,width=0.45 \textwidth]{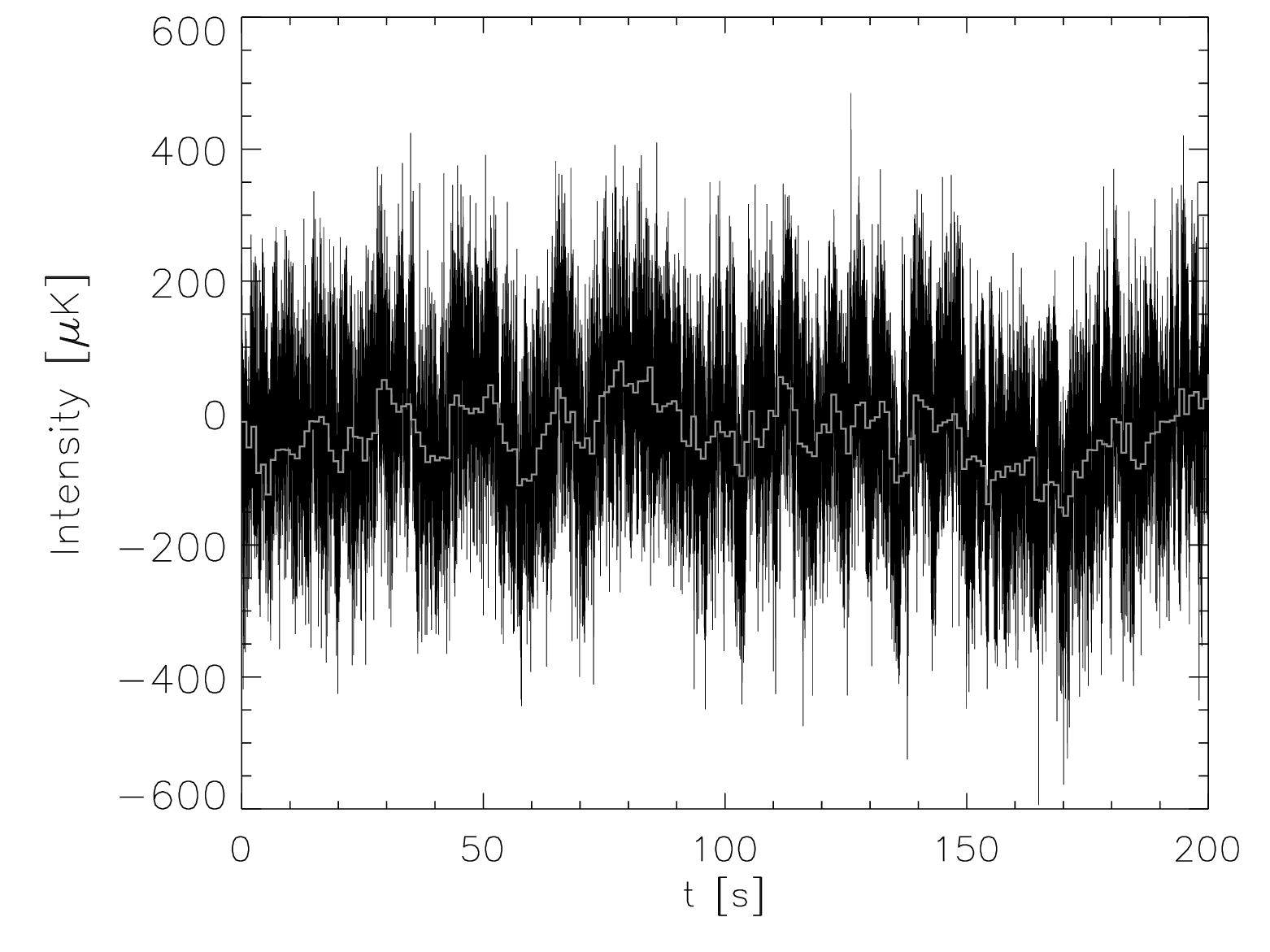}\\%1f_destr.eps}
\includegraphics[angle=0,width=0.45 \textwidth]{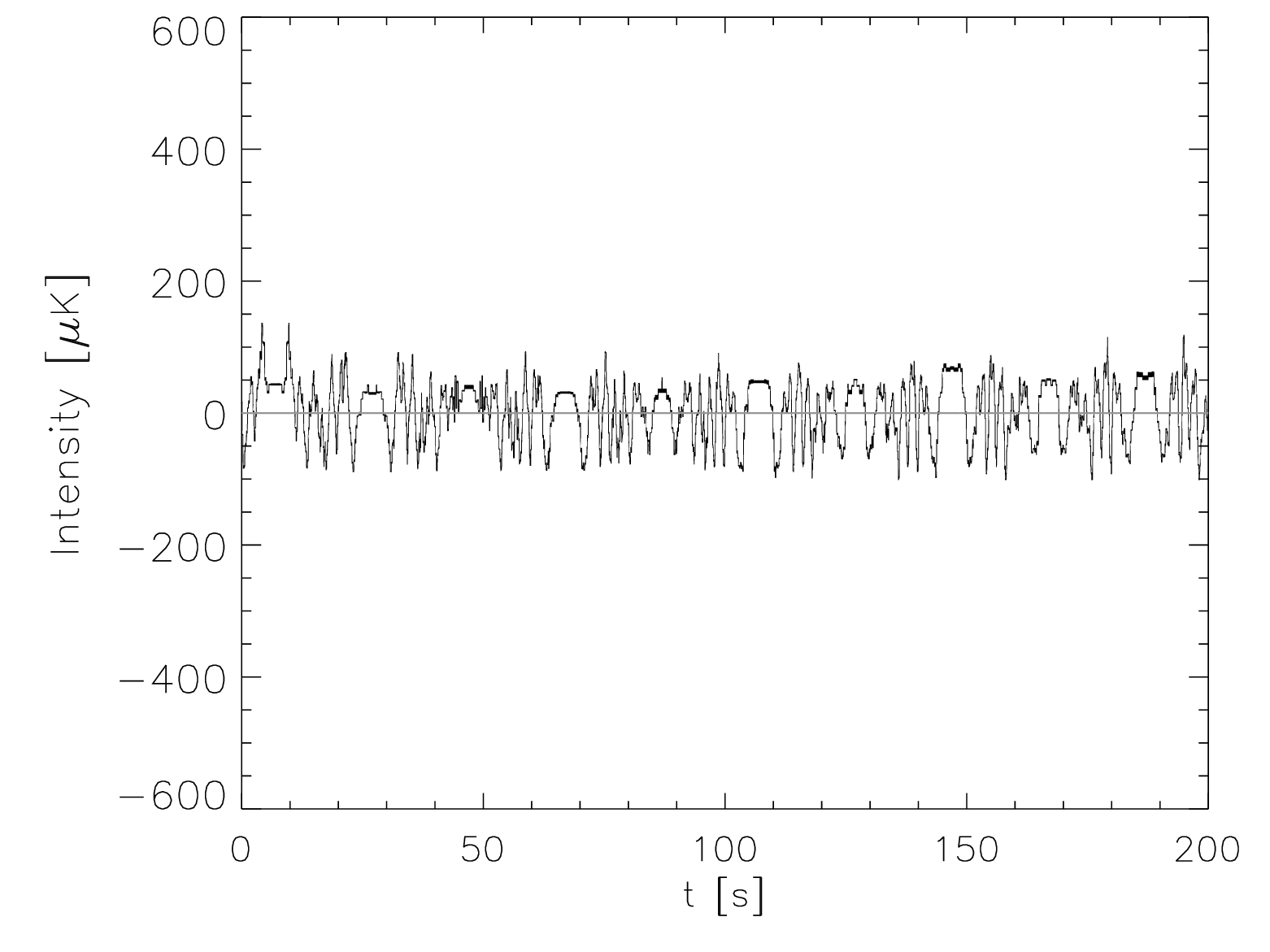}%signal_destr.eps}
\caption{\emph{Upper panel} Plot of a section of raw TOD with long term $1/f$ noise drifts and the maximum likelihood offset functions (light grey) from Descart projected onto it. \emph{Lower panel} Same as above except the TOD is signal only, so the offset amplitudes are all zero.}
\label{1f_destr}
\end{center}
\end{figure}

There is an alternative, approximate, method for modelling the noise
in the time series.  The noise vector $\vec{n}$ can be modelled as uncorrelated white noise plus a series of discrete offsets that represent correlated noise.  The amplitudes of the offsets are estimated and subtracted, as illustrated in Figure \ref{1f_destr}, whilst leaving the signal untouched.  This approach, called destriping, has recently been derived in a maximum-likelihood context (\citealt{keihanen:2004}, \citealt{keihanen:2005}), whose notation we follow.  The latter of these papers describes the MADAM algorithm, whose solution we use here.

We write the noise vector $n_{t}$ as a sum of vectors of uncorrelated (white) noise, $n_{W}$, and correlated noise, $n_{corr}$,
\begin{equation}
n_{t} = n_{w}+n_{corr}
\label{noise dec}.
\end{equation}

The essence of the destriping method is to approximate $n_{corr}$ as an expansion of a set of offset functions, $\mathbf{F_{\alpha}}$, with amplitudes $a$ 
\begin{equation}
n_{corr}= \sum_{i} \mathbf{F_{ti}} a_{i}
\label{corr}.
\end{equation}

The simplest choice of offset function is a constant:

\begin{equation}
\mathbf{F_{ti}} = \left\{ \begin{array}{cc}
1 & \textrm{$t \in \Delta_{i}$} \\
0 & \textrm{otherwise,}
\end{array} \right.
\label{const_chunk}
\end{equation}
where $\Delta_{i}$ is a chunk of the TOD.  In this scheme, $F_{i}a_{i}$ is a discrete jump in TOD space from $0$ outside chunk $i$ to constant $a_{i}$ in side $i$.  Thus, the sum in (\ref{corr}) produces a time domain vector of constant offsets with amplitudes $\vec{a}$ approximating the correlated noise. More complicated offset functions can be chosen, for example Fourier series or Legendre polynomials.

The amplitudes $\vec{a}$ will be Gaussian random numbers satisfying
\begin{eqnarray}
\langle a_{\alpha} \rangle &=& 0 \\
\langle a_{\alpha} a^{T}_{\alpha} \rangle &=& \mathbf{C_{a}}_{\alpha \alpha '}
\label{a Ca constraints}.
\end{eqnarray}

With this approximation for correlated noise, equation (\ref{tod hard}) for the TOD can be re-written as
\begin{equation}
\vec{y}= \mathbf{P} \vec{x} + \mathbf{F} \vec{a} + \vec{n}_{W}
\label{TOD_destr}.
\end{equation}

Again we minimise the $\chi^2$, but this time we require solutions for both the map $\vec{x}$ and the amplitude vector $\vec{a}$.  With $\vec{a}$ as a second parameter, the likelihood of the TOD becomes
\begin{equation}
P(\vec{y}) = P(\vec{y}|\vec{x},\vec{a}) P(\vec{a}|\vec{x}) P(\vec{x})
\label{destr_bayes}.
\end{equation}

However, the probability distribution of the amplitudes is independent of the CMB, which we consider to be deterministic and have no associated probability distribution. Thus, the amplitude probability distribution becomes an independent prior on $\vec{a}$, $P(\vec{a}|\vec{x})= P(\vec{a})$, and the CMB prior is a constant.  The likelihood is now given by

\begin{equation}
P(\vec{y}) = P(\vec{y}|\vec{x},\vec{a}) P(\vec{a}) 
\label{simplified destriping posterior}.
\end{equation}

The likelihood in (\ref{simplified destriping posterior}) is simply the white noise distribution, which is a Gaussian with covariance $\mathbf{N}= \sigma^{2} \delta_{ij}$

\begin{equation}
 P(\vec{y}|\vec{x},\vec{a})= (2\pi \mathbf{|N|})^{-1/2} \exp \bigg(-\frac{1}{2} n_{w}^{T} \mathbf{N^{-1}} n_{w} \bigg)
\label{destriping likelihood}.
\end{equation}

The next step is to decide what prior information on the amplitudes to include through $P(\vec{a})$.  If we have an estimate of the $1/f$ noise power spectrum, we can include prior information through the noise covariance matrix $\mathbf{C_{N}}$.  The probability distribution for the offsets is Gaussian

\begin{equation}
P(\vec{a}) = (2\pi |\mathbf{C_{a}}|)^{-1/2} \exp{\bigg(-\frac{1}{2} \vec{a}^{T} \mathbf{C_{a}^{-1}} \vec{a}\bigg)}
\label{offsets prior},
\end{equation}
where $\mathbf{C_{a}}$ is a re-projection of $\mathbf{C_{N}}$ through $\mathbf{C_{a}}= (\mathbf{F^{T}F})^{-2} \mathbf{F^{T} C_{N} F}$, which for our simulations is circulant (but for real data is symmetric Toeplitz and would be inverted and convolved as described in Section \ref{mle}).  Forming a $\chi^{2}$ from these distributions we obtain a function to minimise for the amplitudes and the map:

\begin{eqnarray}
\chi^{2} & = & -2 \ln| P(\vec{y})|  \nonumber \\
 & = & (\vec{y}- \mathbf{F}\vec{a}-\mathbf{P}\vec{x})^{T} N^{-1} (\vec{y}- \mathbf{F}\vec{a}-\mathbf{P}\vec{x}) \nonumber \\
 & & + \vec{a}^{T} \mathbf{C_{a}^{-1}} \vec{a}
 \label{simple destriping chisq}.
 \end{eqnarray}

The $\chi^{2}$ can be simplified by writing the map $\vec{x}$ in terms of the data through $\vec{x}= (\mathbf{P^{T} N^{-1} P})^{-1} \mathbf{P^{T} N^{-1}} (\vec{y}-\mathbf{F}\vec{a})$.  This equation can be recognised as simple naive binning, as  the noise in the destriped time-stream $\vec{y} - \mathbf{F}\vec{a}$ is white $N= \sigma^{2} \delta_{ij}$.  

We can now gather terms involving $\mathbf{P}$ and $\mathbf{P^{T}}$ into a single operator $\mathbf{Z}$

\begin{equation}
\chi^{2} = (\vec{y} - \mathbf{F} \vec{a})^{T} \mathbf{Z^{T} N^{-1} Z} (\vec{y} - \mathbf{F}\vec{a})+ \vec{a}^{T} \mathbf{C_{a}^{-1}} \vec{a}
\label{simple destriping chisq with Z}, 
\end{equation}
where
\begin{equation}
\mathbf{Z} = \mathbf{I} - \mathbf{P(P^{T} N^{-1} P)^{-1} P^{T} N^{-1}}
\label{definition of Z}
\end{equation}
and $\mathbf{I}$ is the identity matrix.

The operator $\mathbf{Z}$ is a signal cleaning operator.  The operation $\mathbf{Z} \vec{y}$ removes the signal component from $\vec{y}$ by subtracting a naive map of $\vec{y}$ projected onto TOD space.  $\mathbf{Z}$ also has the property $\mathbf{Z^{T} N^{-1} Z} = \mathbf{N^{-1} Z}$.

We obtain an estimator for $\vec{a}$ by minimising (\ref{simple destriping chisq with Z}) with respect to $\vec{a}$

\begin{equation}
(\mathbf{F^{T}N^{-1}ZF + C_{a}^{-1}})\vec{a} = \mathbf{F^{T} N^{-1}Z}\vec{y} 
\label{amp_destr}.
\end{equation}

This is an inverse problem like that of the maximum likelihood algorithm.  We can use many of the same tricks in solving it.  We solve the system iteratively, using a preconditioned conjugate gradients (pcg) algorithm.  The system is typically smaller than that in the previous section because $N_{chunks} < N_{pixel}$ for a single day.  Again, we do not explicitly construct the matrix, rather we do each operation on both sides of (\ref{amp_destr}) individually.

This system solves much more quickly than the maximum likelihood system.  The operations on the left hand side of (\ref{amp_destr}) scale as $\mathcal{O}(N_{t})$. The exception to this is the inversion of the circulant offset covariance matrix $\mathbf{Ca}$, which can be achieved through cyclic deconvolution scaling as $O(n_{a} \log_{2}|n_{a}|)$. Each iteration of the destriping algorithm scales as $\mathcal{O}(N_{t} + n_{a} \log_{2}|n_{a}|)$, in comparison to $\mathcal{O}(N_{t}(1+ \log_{2}|N_{t}|))$, the iterative scaling of the maximum likelihood algorithm.

The system is preconditioned using preconditioner $K$ such that $K^{-1} A \vec{x} = K^{-1} \vec{b}$, where 

\begin{equation}
K = \mathbf{F^{T} N^{-1} F} + \mathbf{C_{a}^{-1}}
\label{Descart preconditioning}
\end{equation}
is a circulant matrix and is inverted using cyclic deconvolution.

With the amplitude vector found, one subtracts the correlated noise approximation $\mathbf{F}\vec{a}$ from the TOD $\vec{y}$ and naively bins the cleaned TOD to return the destriped map

\begin{equation}
\vec{x} = (\mathbf{P^{T} P})^{-1} \mathbf{P^{T}} (\vec{y}-\mathbf{F}\vec{a})
\label{destr_map},
\end{equation}

The noise covariance of this map will be 

\begin{equation}
\mathbf{C}= (\mathbf{P^{T}} (\mathbf{N} + \mathbf{F C_{a} F^{T}})^{-1} \mathbf{P})^{-1}
\label{destriped map covariance},
\end{equation}
where $\mathbf{N} = \langle n_{w} n_{w}^{T} \rangle = \sigma^{2} \delta_{ij}$ is the white noise covariance and $\mathbf{C_{a}} = \langle a a^{T} \rangle$ is the covariance of the amplitudes.

\section{Simulations} 
\label{simulations}

We produced 4 sets of simulations, each with 200 signal+noise and 200 further noise only realisations, from 12 hours of observing using a single detector sampling at $f_{sam}= 100$-Hz for 2 scanning strategies.  The simulations are summarised in Table \ref{simulation table}. The experimental design was that of a modulation experiment scanning at $1^{o} s^{-1}$ with a half wave plate rotating at $f_{rot}= 5$-Hz, corresponding to a polarisation modulation frequency $f_{mod}= 20$-Hz.

Our aim is to examine the capabilities of different algorithms on the same data.  For this goal, the white noise variance is essentially irrelevant: what we care about is the level of correlation between TOD and thence map pixels.  To see B-modes will require years of observing with hundreds of detectors.  For this work, in which we are simulating 12 hours of data from a single detector,  we have chosen a white noise level that allows us to convincingly measure B-modes.   For all our simulations, we use a heuristic NET$= 0.242$ $\mu K \sqrt{s}$.

Two noise scenarios are simulated.  The first of these represents detector dominated $1/f$, as would be returned for a space-bourne or balloon flight experiment (like MaxiPol \citep{johnson:2007} or Archeops \citep{macias-perez:2007}) and has a $1/f$ power spectrum with spectral index $\alpha= 1.0$ and knee frequency $0.1$-Hz, the projected noise correlation properties of the C$_{l}$OVER detectors. 

The second scenario simulates $1/f$ dominated by atmospheric fluctuations, assumed to be un-polarised, which is expected to be the case for ground based experiments. The atmospheric $1/f$ noise is simulated with spectral index $\alpha=1.9$ and knee frequency $f_{knee}=0.2$-Hz . We calculate these numbers from a rough fit to sample QUaD 100 GHz noise power spectra (see figure 36 of \citealt{hinderks:2008}) and they are fiducial: the level of atmospheric fluctuation depends upon wind and scanning velocities and evidence suggests it may vary between CMB observing sites \citep{bussman:2005}.

%evidence suggests that atmospheric fluctuation is greater at the Atacama CMB sites than at the South Pole \citep{bussman:2005} and the atmospheric noise.

\begin{table}
\caption{Summary of the simulations}
\begin{center}
\begin{tabular}{c|c|c|c}
Strategy& tensor to scalar ratio & knee frequency & spectral index\\
\hline
Sabre & $r=0.1$ & $f_{knee}=0.1$-Hz & $\alpha=1.0$ \\
%\hline
Sabre & $r=0.1$ & $f_{knee}=0.2$-Hz & $\alpha=1.9$ \\
%\hline
Sabre & $r=0.0$ & $f_{knee}=0.1$-Hz & $\alpha=1.0$ \\
%\hline
Fence & $r=0.1$ & $f_{knee}=0.1$-Hz & $\alpha=1.0$ \\
\hline
\end{tabular}
\end{center}
\label{simulation table}
\end{table}%

The TOD is simulated by taking a list of telescope pointings from a scanning strategy and pulling out the T, Q and U signals from simulated CMB maps at the right ascension and declination of the pointing.  The T, Q and U signals are combined into time streams using equation (\ref{mod_timestream}) and added to an instrumental noise stream.

\subsection{CMB template}
\begin{figure}
\begin{center}
\includegraphics[angle=0,width=0.45 \textwidth]{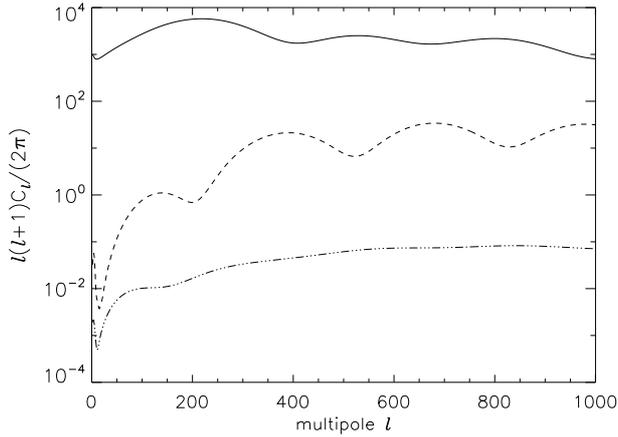}
\caption{Input theoretical power spectra: the solid curve is $C_{l}^{TT}$; the dashed curve is $C_{l}^{EE}$; and the dot-dashed curve is $C_{l}^{BB}$}%lower panel is combined beam and pixel window function}
\label{cmb template}
\end{center}
\end{figure}

Two theoretical power spectra were calculated using the CAMB package \footnote{http://camb.info} \citep{lewis:2000} for a typical set of concordance parameters.  The first spectrum was used to synthesise maps with B mode corresponding to an initial tensor to scalar ratio of $r=0.1$ plus the expected B-modes from weak lensing.  The T, E and B spectra from this model are shown in Figure \ref{cmb template}.  The second spectrum was calculated to produce zero B-mode simulations, using $r=0$ and ignoring weak lensing B-modes.

The model power spectra were convolved with a fiducial symmetric Gaussian instrumental beam of FWHM 10 arc-minutes.  From the initial spectra, 200 Gaussian realisations of the CMB were simulated using the HEALPix \citep{gorski:2005} package \emph{synfast}\footnote{http://healpix.jpl.nasa.gov/html/facilitiesnode11.htm} using $n_{side}=512$, giving resolution up to $l= 1024$.

\subsection{Noise Simulation}

The simulated noise stream is a sum of stationary Gaussian realisations of uncorrelated white noise and correlated $1/f$ noise, which together have power spectral density

\begin{equation}
P(f) = \frac{\sigma^{2}}{f_{sam}} \bigg( 1 + \Big(\frac{f_{knee}}{f}\Big)^{\alpha} \bigg)
\label{noise_f_spectrum},
\end{equation}
where $\sigma^{2}$ is the white noise variance of a single observation, $f_{sam}$ is the \emph{sampling frequency} of the detectors (where integration time $t= 1/f_{sam}$), $\alpha$ is the spectral index and $f_{knee}$ is the knee frequency of the spectrum.

The noise we have simulated is both stationary and Gaussian, satisfying
\begin{eqnarray}
\langle n \rangle&=& 0\\
\langle n n^{T} \rangle &=& \mathbf{C_{N}}
\label{noise mean and cov}.
\end{eqnarray}

Due to the stationarity property, the covariance of the noise between TOD becomes a function of time separation. The noise covariance matrix $\mathbf{C_{N}}$ is circulant and symmetric, allowing us to quickly evaluate its inverse  using a Fourier transform:

\begin{equation}
C_{N}^{-1} (t-t')= \bigg( \frac{2}{N_{t}} \bigg)^{2} \int_{-\infty}^{\infty} P^{-1}(f) e^{-if(t-t')} df
\label{noise_filter},
\end{equation}
where $P^{-1}(f)$ is the inverse noise power spectrum.

\subsection{Scanning strategies}
\label{strategies}

\begin{figure}
\begin{center}
\includegraphics[angle=270,width=0.45 \textwidth]{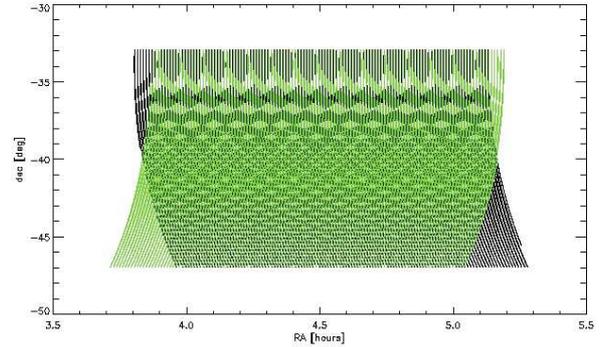}
\caption{Ra/dec plot of the sabre scan pointing. The black curve is the rising field scan and the grey curve is the setting field scan.}
\label{atascan pic}
\end{center}
\end{figure}

We simulate TOD using two scanning strategies.  The noise realisation in the TOD are the same for each scan, so this amounts to changing the signal in TOD.  Scanning strategy has been shown to have an effect on the ability of map-makers to reduce noise correlation \citep{tegmark:1997b}.  Most important is the \emph{connectivity} of the scan (or its degree of \emph{cross-linking}).  With greater cross-linking, map-makers should produce fewer stripes and smaller residuals.  Whilst the isotropy of the scans is different, we have kept the same average integration time per pixel in each map.

Our scanning strategies are the following:

\begin{center}
\begin{enumerate}
\item \emph{Sabre scan}: Simulated scanning strategy for ground based experiment sited in the Atacama desert.  The scanning elevation is kept constant at $45^{\circ} $  to minimise atmospheric noise.  The telescope scans back and forth in azimuth with a sinusoidal velocity curve, whilst the field rises or sets through the scanning elevation.  The field is observed twice per day as it rises and sets, resulting in a minimally cross-linked scan as shown in Figure \ref{atascan pic}.  The experiment runs for $\approx 5$ days to complete 12 hours of integration time. \\
\item \emph{Fence scan}: For half the scanning time, the telescope slews back and forth horizontally with sinusoidal velocity whilst the field moves vertically through the scan at a much slower constant velocity. The scan is repeated for the second half of the scanning time with the scanning direction and field drift directions swapped, producing a highly cross linked square field.  In total, 12 hours integration is completed.
\end{enumerate}
\end{center}

The sabre scan is representative of the typical level of cross-linking achievable from the ground when keeping the scanning elevation constant.  This is a vital constraint, as if the scanning elevation is varied then noise sourced from changes in airmass swamps the faint CMB signal.  Unfortunately, realistic scans are never ideally cross-linked. The fence scan represents a nearly ideally cross-linked scan that cannot be achieved from the ground, but it is included to probe the effects of cross-linking on the performance of destriping.

\subsection{Diagnostics}
\label{diagnostics section}

We require from our diagnostics, comparable statistics indicating the fitness for purpose of the map-making algorithms.  We use the following diagnostics:

\begin{enumerate}
\item \emph{RMS residual}: Root-mean-square of the residual map between the recovered maps and the input theoretical map used to generate the TOD. 
\item \emph{Residual angular power spectrum}: spatial information about the comparative temperature residuals is gained by analysing the angular power spectrum of the residual maps.  The residual map is convolved with map field's window function $W_{l}$, so a modification of the MASTER method \citep{hivon:2002} is used to return an unbiased binned estimate of the residual (noise) power spectrum.
\item \emph{Pure pseudo-$C_{l}$ E and B mode estimation}: The pure pseudo-$C_{l}$ estimator (\citealt{ksmith:2006}, \citealt{ksmith:2007}) is used to return estimates of the E and B mode angular power spectra $C_{l}^{EE}$ and $C_{l}^{BB}$ from the estimated Q and U maps.  This estimator has been shown \citep{ksmith:2006} to return unbiased estimates of E and B without the E$\rightarrow$B mixing from the ambiguous modes that arise from the field boundaries \citep{bunn:2002}.  The diagnostics are the mean estimates of the power spectra, the mean estimates of the E and B noise power spectra and the Monte-Carlo error bars for a single realisation.  The details of the pure pseudo-$C_{l}$ estimator are presented in Appendix \ref{pseudo Cl appendix}.
\item \emph{Filter function}: The filter function $F_{l}$ from a preliminary signal only simulation is used to analyse the signal error component to the estimated map residuals. The filter transfer function $F_{l}$ \citep{hivon:2002} gives the degree to which signal filtering by the map-making process affects the recovered power spectrum.  If we don't want to \emph{lose} any information on the signal then the imprint of filtering must be negligible ($F_{l}-1 = 0$).

The power spectrum of the estimated map $\tilde{C_{l}}$ is related to that of the input map $C_{l}^{\mathbf{I}}$ by (\cite{natoli:2001}, \cite{poutanen:2006})

\begin{equation}
\langle \tilde{C_{l}} \rangle = F_{l} \langle C_{l}^{\mathbf{I}} \rangle + \langle N_{l} \rangle 
\label{filter function equation}
\end{equation}
where $N_{l}$ is the power spectrum of the noise bias. If the estimated map is made from signal only TOD, $\langle N_{l} \rangle =0$ so $F_{l}$ can be calculated by inverting (\ref{filter function equation}).

Both $\tilde{C_{l}}$ and $C_{l}^{\mathbf{I}}$ are convolved with the same beam and have the same pixelisation and sky mask. They have identical transfer matrices $K_{ll'}$ (see Appendix \ref{pseudo Cl appendix}), so we need not correct for mode coupling. 
\end{enumerate}

\section{Results} 
\label{results}

The simulations were analysed by three map-making algorithms: standard naive map-making (equation \ref{naive estimator}), optimal maximum likelihood map-making, and destriping.  Destriping was repeated with various offset function baseline lengths (hereafter denoted by $\lambda_{C}$ in units of time), ranging from a baseline length  $\lambda_{C}=1$ second through to $\lambda_{C}=1000$ seconds ($\approx 16$ minutes).  The destriping code, Descart, operated in two modes: \emph{traditional} destriping mode (as in all papers prior to \cite{keihanen:2005}), and in a \emph{covariant} destriping mode, making use of noise information through the offset covariance matrix $C_{a}$ \citep{keihanen:2005}.  The two modes are compared in Section \ref{Ca section}. Elsewhere, all destriping is the covariant form.

Both the covariant destriping and maximum likelihood map-makers require prior noise information (through  $C_{a}$ and $C_{N}^{-1}$ respectively).  In reality, this noise information would have to be estimated from the TOD directly, highlighting the importance of the noise estimation step immediately prior to map-making.  In this paper, we are not addressing the issues of noise estimation and so use ensemble average noise information through the power spectrum $P(f)$ used to generate the TOD noise.  A number of approaches to noise estimation have been discussed in the literature (eg: \citealt{ferreira:jaffe:2000}, \citealt{amblard:hamilton:2004}).

\subsection{Signal only maps} 
\label{signal}

\begin{figure}
\begin{center}
\includegraphics[angle=0,width=0.48 \textwidth]{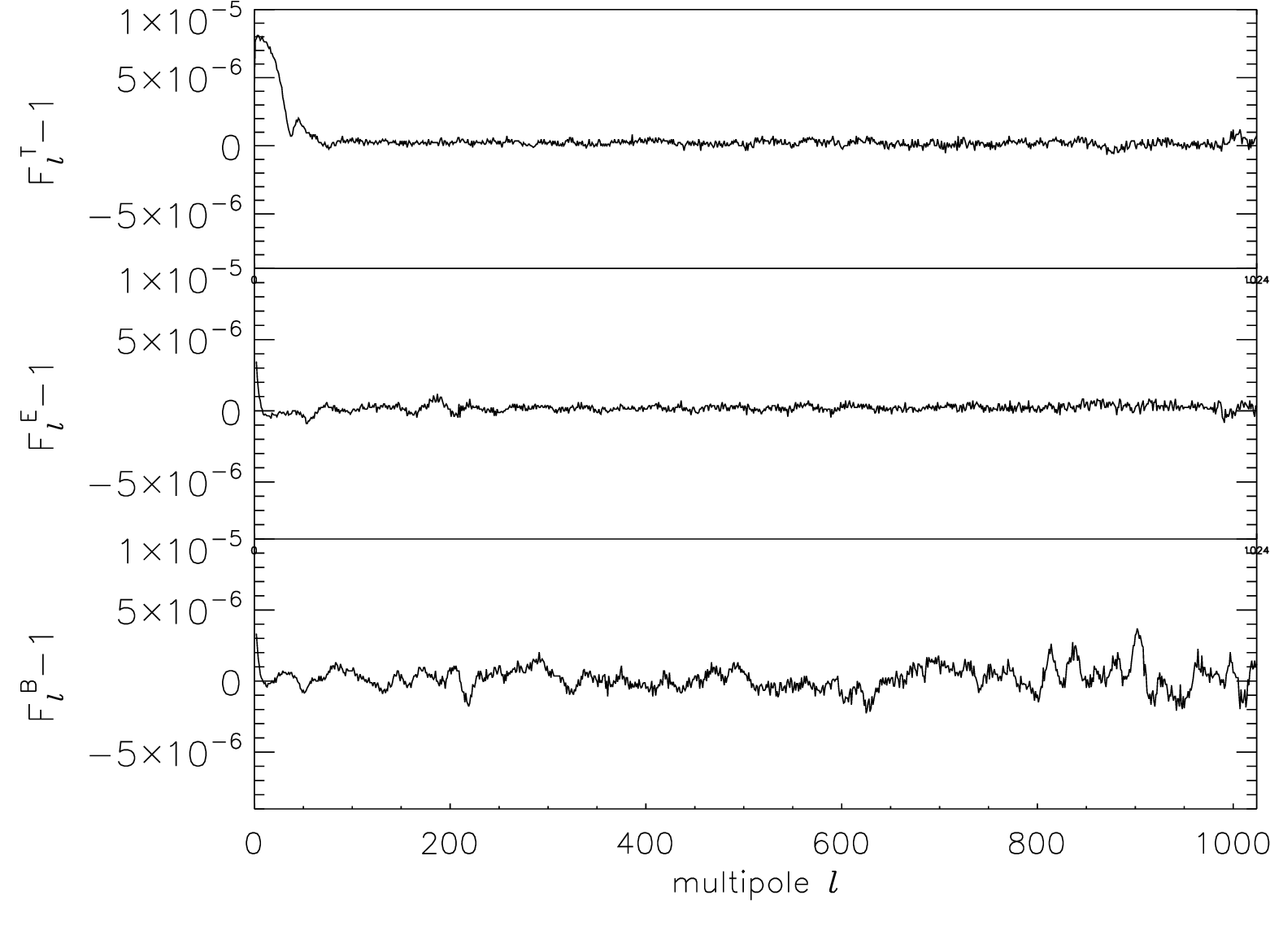}\\%{plots/mle_t_filterfn.eps}
\includegraphics[angle=0,width=0.48 \textwidth]{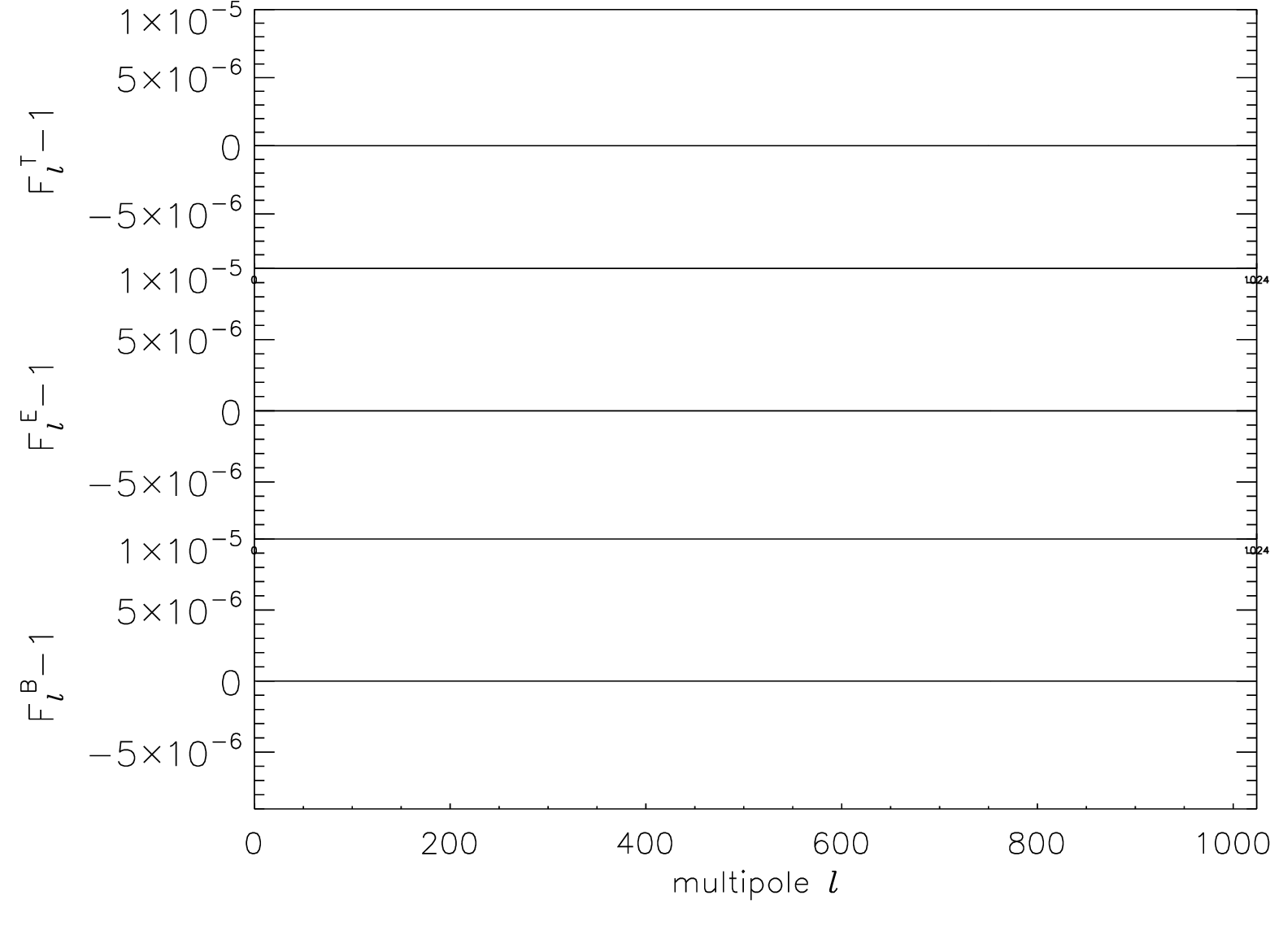}\\%{plots/destr_filterfn.eps}
\includegraphics[angle=0,width=0.48 \textwidth]{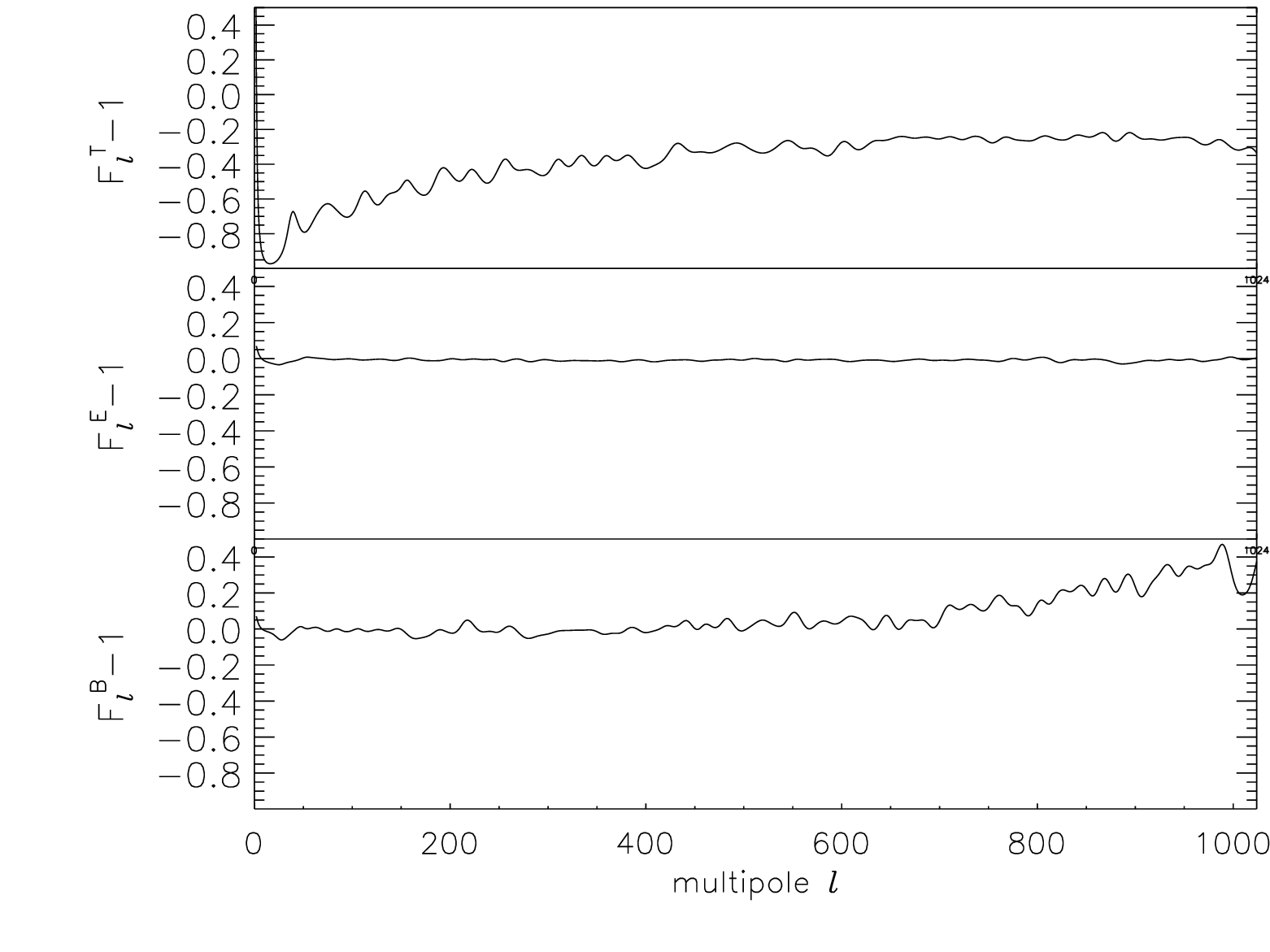}%{plots/MLE_T_sigerror.eps}
\caption{\emph{Upper panel} Filter function $F_{l}-1$, for maximum likelihood algorithm showing minimal signal degradation from the initial \emph{signal only} simulation. \emph{Middle panel} $F_{l}-1$ for destriping, displaying zero signal degradation. \emph{Lower panel} Example filter function for naive map of TOD filtered with $\tilde{C_{N}}^{-1/2}$ filter, where the signals are distorted and T is heavily filtered}
\label{T_filter_fn}
\end{center}
\end{figure}

In addition to signal+noise simulations, an initial pure signal simulation was analysed first for the sabre strategy, using the same prior noise information as for the signal+noise simulations. In this limit, any departures in the estimated map from the input map are distortions of the signal caused by the map-making process itself.  The filter functions for this simulation were calculated using (\ref{filter function equation}) where $C_{l}^{I}$ and $\hat{C_{l}}$ are raw pseudo-$C_{l}$s calculated by the HEALPix package \emph{anafast}\footnote{http://healpix.jpl.nasa.gov/html/facilitiesnode5.htm} \citep{gorski:2005}.

Figure \ref{T_filter_fn} shows the T, E and B filter functions for the maximum-likelihood (upper panel) and destriping (middle panel) algorithms.  Destriping, covariant or otherwise, does not filter the signal at all, despite use of a noise prior through $C_{a}$ (Figure \ref{T_filter_fn}, middle panel).  The data term in the likelihood (\ref{simple destriping chisq with Z}) forces the offset amplitude vector to the null vector, as the signal cleaning operator $\mathbf{Z}$ perfectly removes the signal from the fit.

The maximum likelihood algorithm does display some minimal signal distortion (Figure \ref{T_filter_fn}, upper panel).  The magnitude of the distortion is effectively negligible, amounting to less than $0.001\%$ of the signal at the worst multipoles ($l < 50$ for the T map).  The filtering satisfies $\sigma_{filtering} << \sigma_{noise}$ at all multipoles and can be ignored.

This filtering can be explained through the presence of degenerate pixels.  A minimum of three observations at different modulation angles $\beta$ are required to reconstruct T, Q and U for each pixel.  Any pixel for which this condition is not met is degenerate and must be ignored, else the problem becomes singular.  

For Descart, the reconstruction inversion is conducted for each pixel separately - if any pixel's $3\times3$ pointing matrix is singular, then the pixel is irrecoverable.  For the maximum-likelihood algorithm, the reconstruction is accomplished by the matrix inverted by the MINRES conjugate gradient inverter.  If degenerate pixels are included, the MINRES iterations do not converge: the improvement in residuals per iteration tends to zero before a reasonable convergence critereon is reached (such as $10^{-6}$, \citealt{patanchon:2007}).  If near-degenerate pixels remain, the condition number of the matrix increases, requiring more iterations to invert the matrix (see Table \ref{sig conv table}).

\begin{table}
\caption{Comparison of the performance of iterative inversion in the maximum-likelihood algorithm for different pixel exclusion conditions in the signal only simulation. $N_{min}$ is the minimum number of hits required to accept a pixel whilst convergence is the rms residual upon exit of the MINRES algorithm. The $N_{min}=2$ case includes degenerate pixels, so the matrix is singular and the iterations fail before the convergence critereon is reached.}
\begin{center}
\begin{tabular}{c|c|c|c}
$N_{min}$ & \#iterations & $N_{pixel}$ & convergence \\
\hline
$13$ & $90$ & $17499$ & $10^{-6}$  \\
$5$ & $90$ & $17520$ & $10^{-6}$ \\
$4$ & $100$ & $17524$ & $10^{-6}$ \\
$3$ & $101$ & $17531$ & $10^{-6}$ \\
$2$ & $87^{*}$ & $17536$ & $7.5\time 10^{-4}$ \\
\hline
$^{*}$ iterations failed&&&
\end{tabular}
\end{center}
\label{sig conv table}
\end{table}%

Such pixels will be common for ground based polarisation experiements, where the TOD is naturally split into day chunks and maps are made daily.  It is well established that these pixels must be removed in such a way as to maintain time stream continuity \citep{stompor:2002}.  Time ordered data sourced from a degenerate pixel is referenced to a junk pixel with zero signal outside of the estimated map that is ignored in the minimisation.  Degenerate TOD are kept for the noise deconvolution step $\mathbf{C_{N}^{-1}}\vec{y}$, where the TOD value is replaced by a constrained realisation of noise.

However, some errors from this method propagate through the deconvolution step of (\ref{prewhiting matrix mult}), which is applied approximately through FFTs, into neighbouring pixels along the scan direction, causing signal filtering.

This effect does not appear in the Descart destriped maps, as the degenerate TOD are removed from the  binning/projection process of $\mathbf{F^{T}}$ and $\mathbf{F}$ respectively.  There are no resultant gaps in the offset amplitude vector $\vec{a}$ and so the evaluation of $\mathbf{C_{a}^{-1}} \vec{a}$ using FFTs suffers no degradation due to discontinuities. 

For comparison, the lower panel Figure \ref{T_filter_fn} shows the filter function for high-pass filtered TOD using the $C_{N}^{-1/2}$ filter.  In Fourier space, this filter is the square root of the noise filter in (\ref{noise_filter}).  It has the property that filtered TOD with $1/f$ noise have the same diagonal noise covariance matrix as TOD with white noise only, pre-whitening the TOD.  The effect of the filter on the T signal is devastating.  The effect on Q and U is mitigated here due to the modulation.  Whilst the mean filter functions can be evaluated from Monte-Carlo simulations and then deconvolved from the power spectrum \citep{hivon:2002},  the wiggles in the functions are signal realisation dependent and will add to the variance of the recovered power spectrum.  We suspect the high-$\ell$ bias in the B-mode filter function is due E$\rightarrow$B mixing caused by distortion of the Q and U signals by the filtering.  This will be studied in the next paper in the series.

The filtering effects here are separate from the signal filtering reported in \cite{poutanen:2006}, \cite{ashdown:2007a} and \cite{ashdown:2007b}.  Our input CMB maps are at the same resolution (HEALPix $n_{side}=512$) as the estimated maps, so the filtering error from applying FFTs to TOD including sub-pixel signal gradients is absent.  We have considered the sky to be innately pixelised at the experimental resolution, which is one of the underlying assumptions in map-making formalism.  If the sky were pixelised as assumed, destriping would be lossless to machine precision.

Including sub-pixel gradients, it is expected that the maximum likelihood algorithm would filter the signal more than reported here: this is an avenue of on-going research.

\begin{figure*}
\begin{center}
\includegraphics[angle=0,width=0.33 \textwidth]{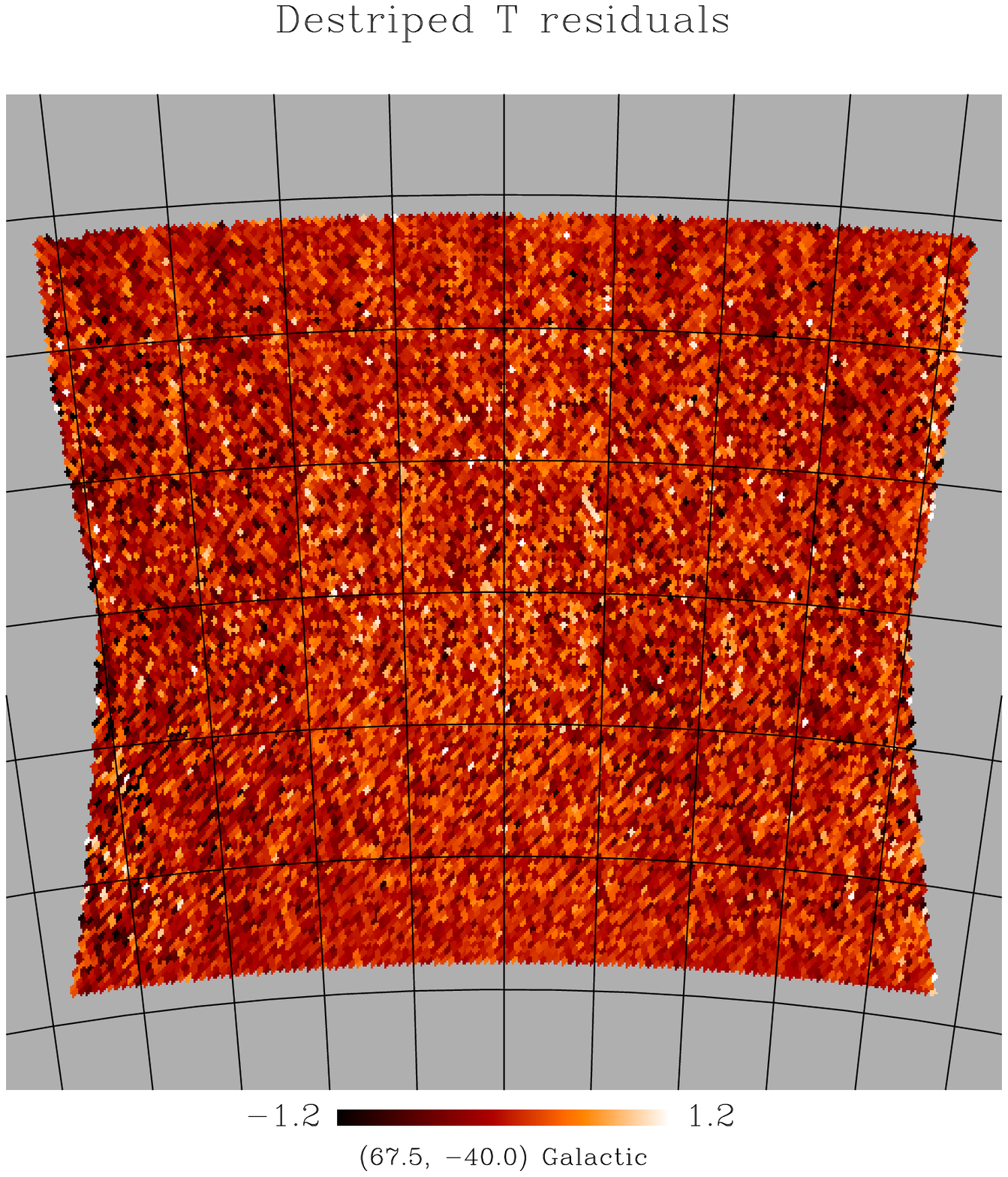}% t destriped residuals
\includegraphics[angle=0,width=0.33 \textwidth]{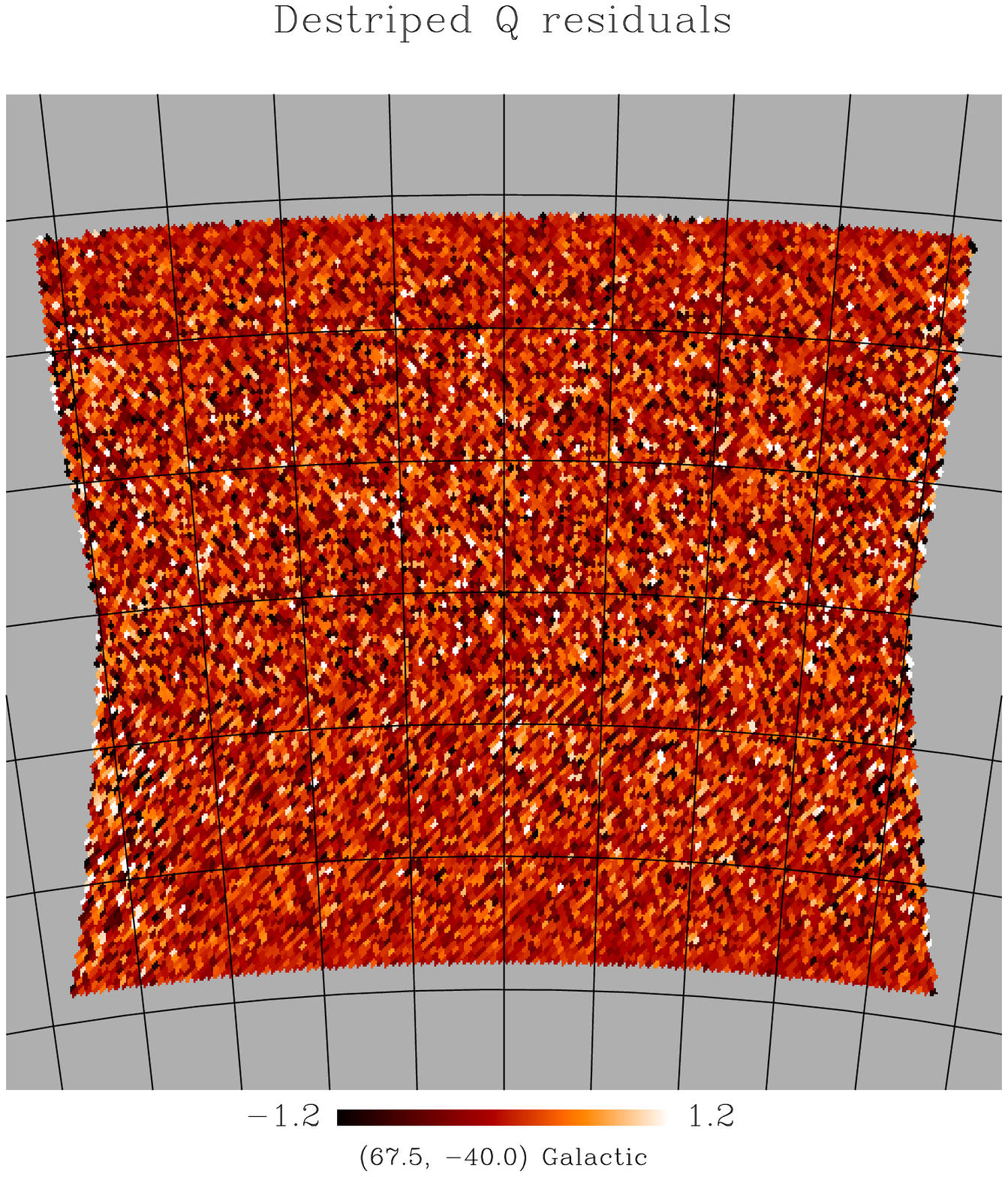}% q destriped residuals
\\
\includegraphics[angle=0,width=0.33 \textwidth]{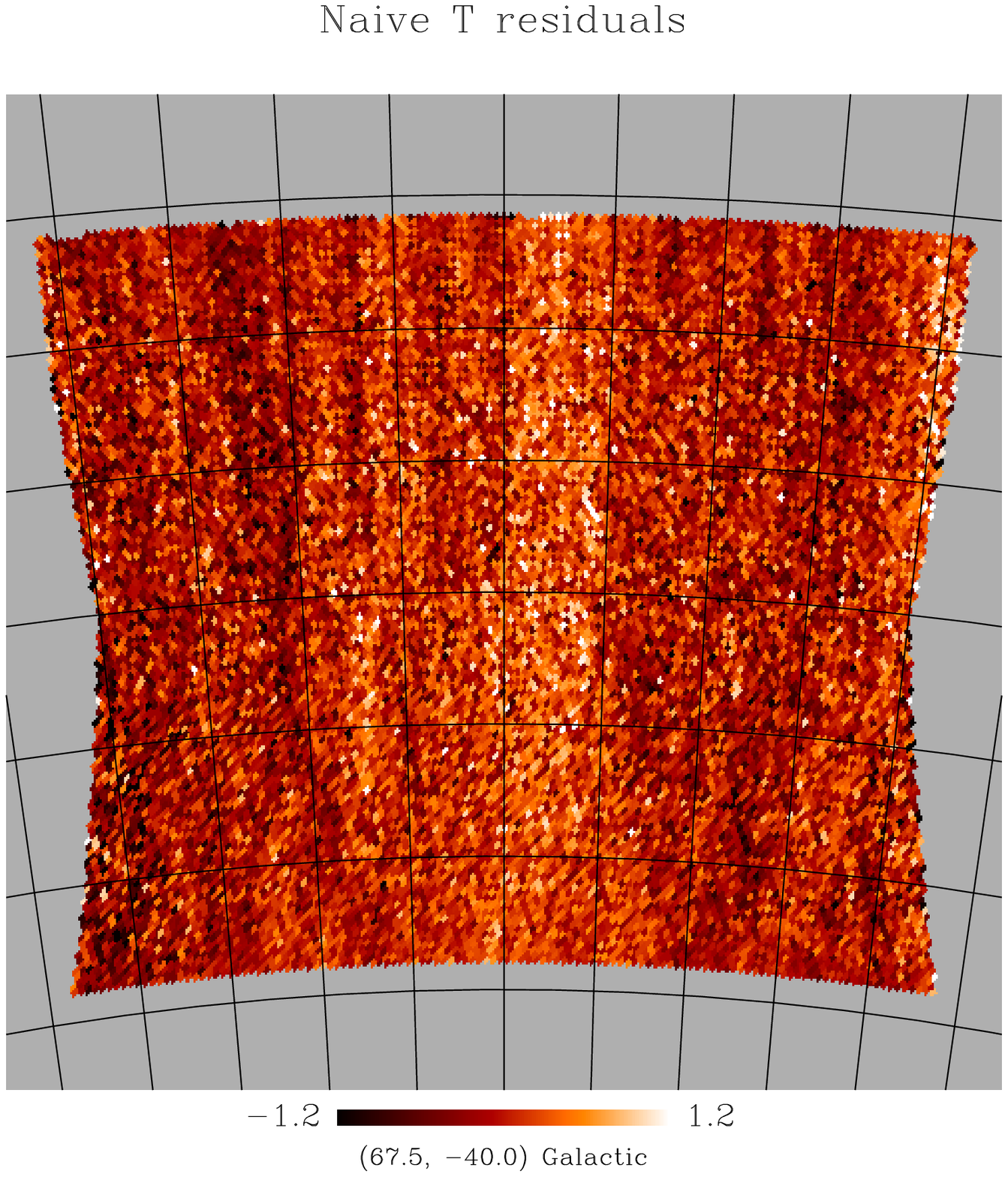}
\includegraphics[angle=0,width=0.33 \textwidth]{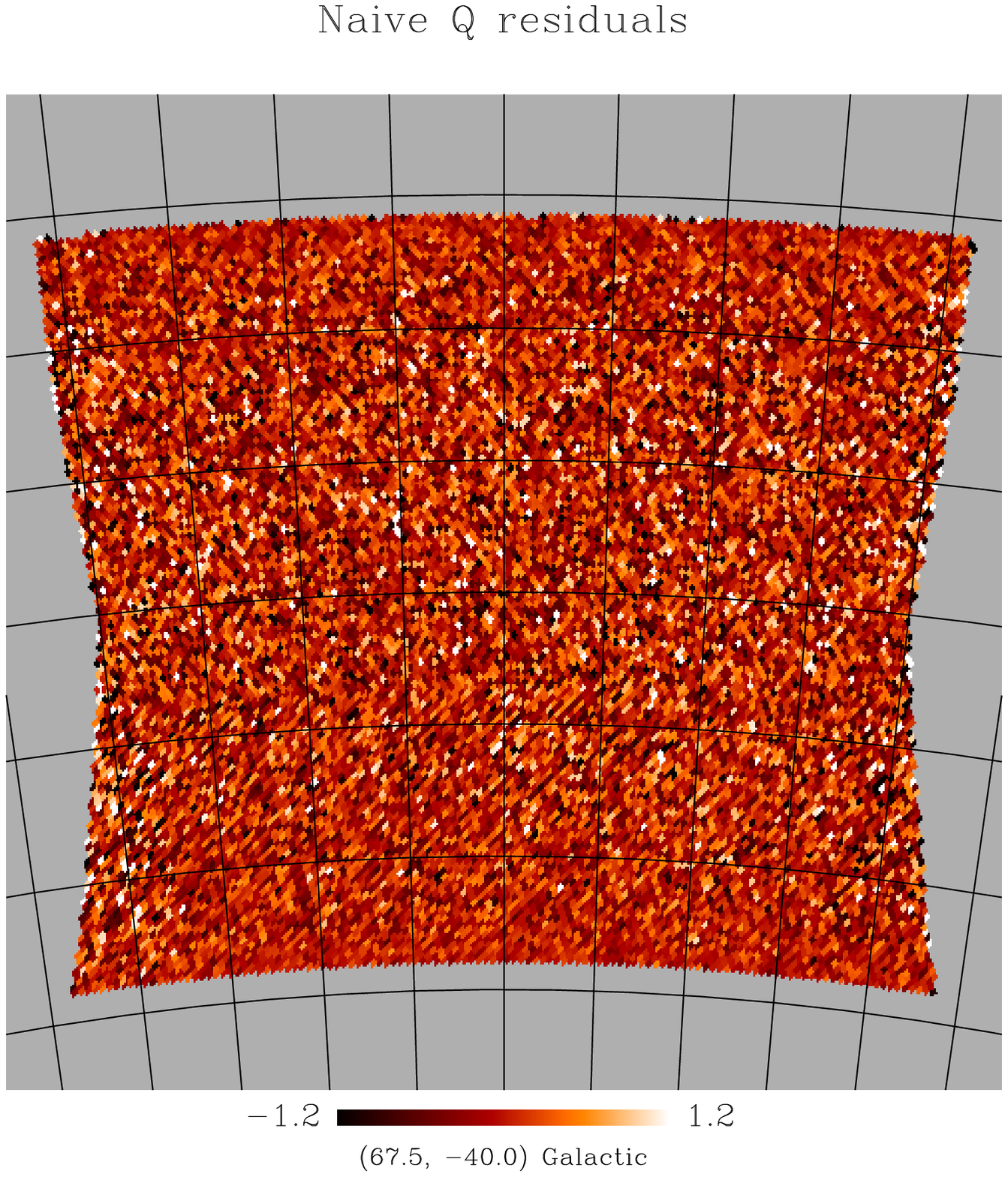}
\\
\includegraphics[angle=0,width=0.33 \textwidth]{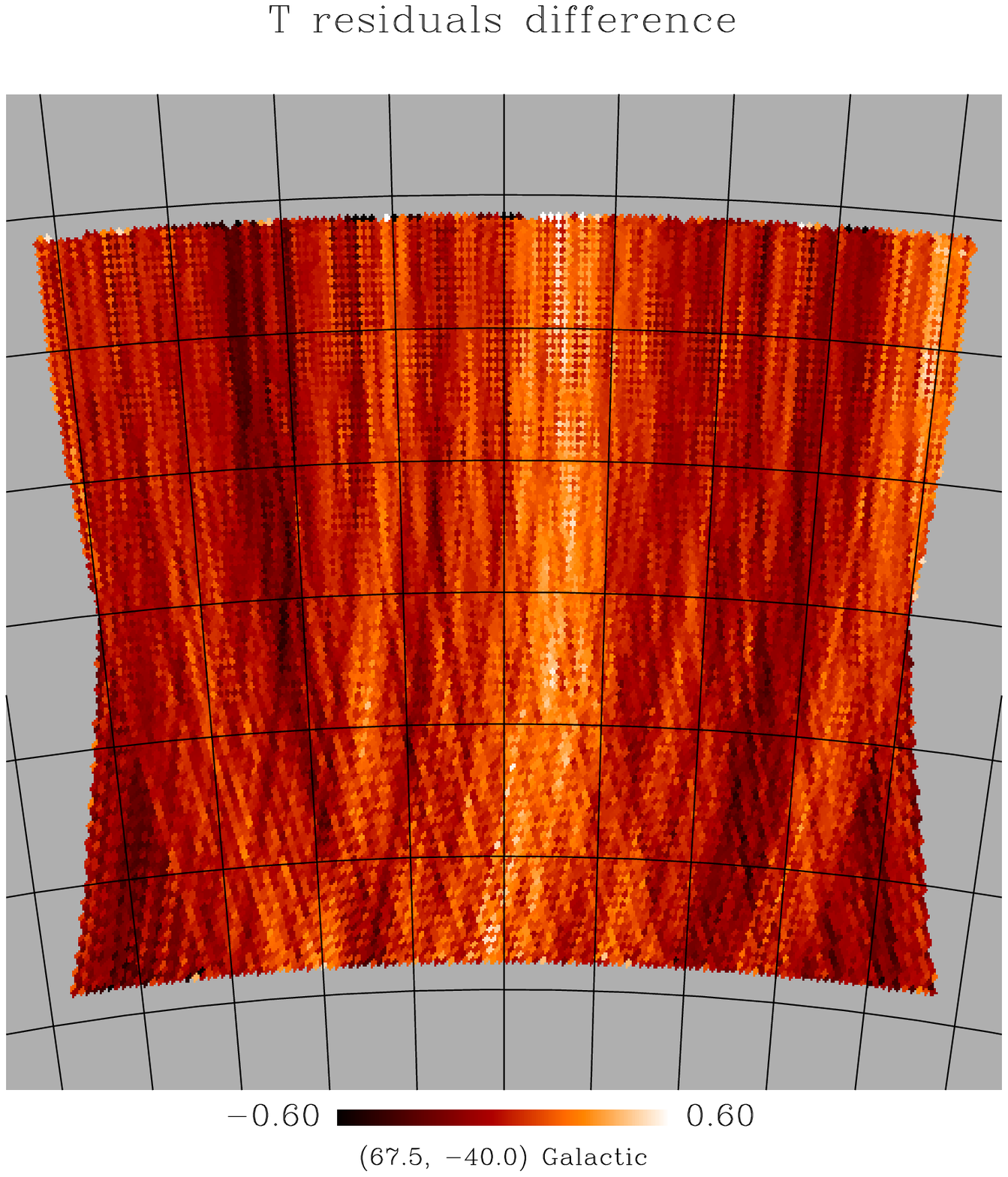}% naive- destriped
\includegraphics[angle=0,width=0.33 \textwidth]{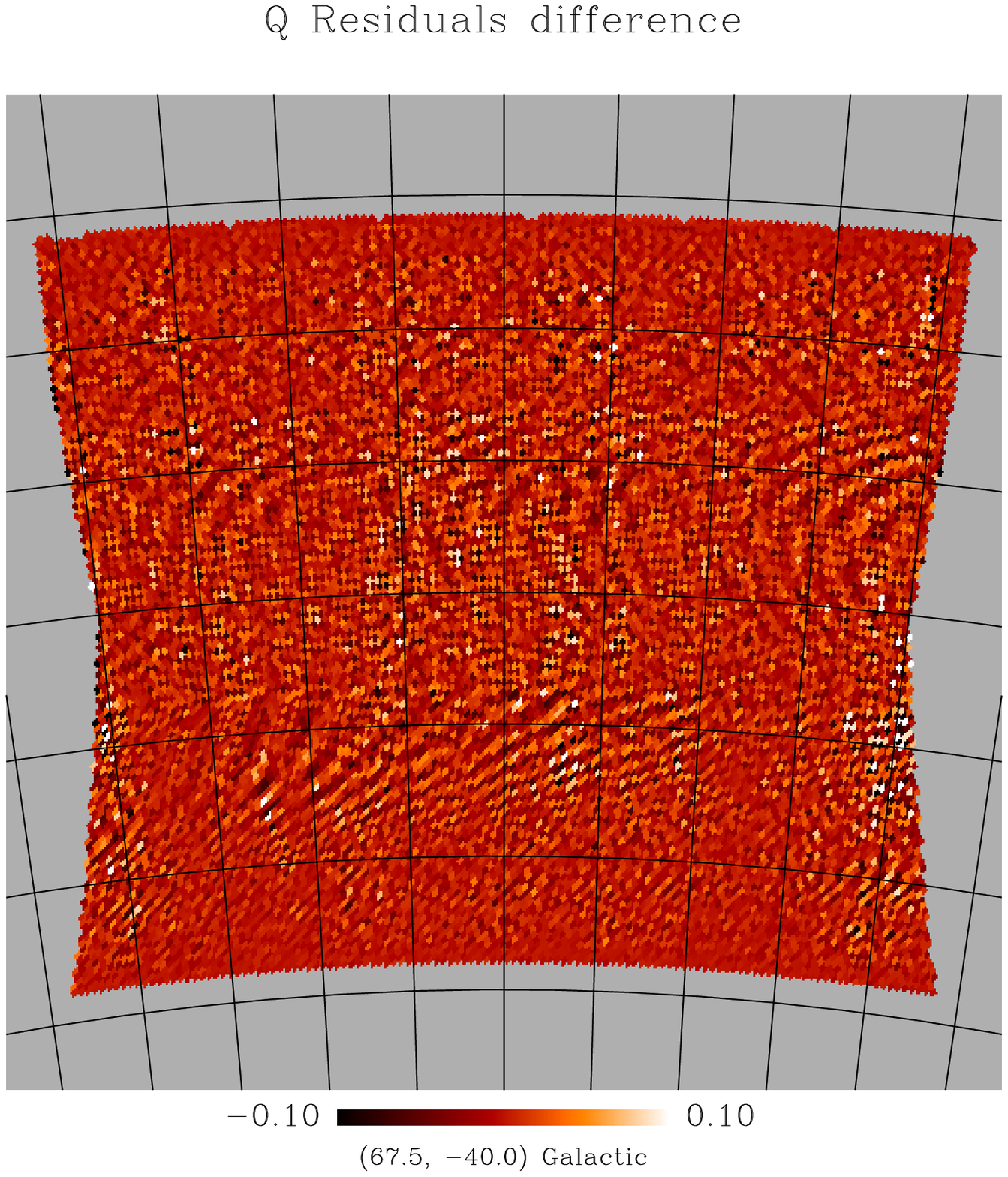}% naive-destriped
\\
\caption{\emph{Top row}: Residuals between a Descart $1-$s output map and the input map, T on the left and Q on the right.
\emph{Middle row}: Residuals between a naive output map and the input map, T on the left and Q on the right.
\emph{Bottom row}: Difference between the top an middle row: Naive residuals - Descart residuals.  Again, T is on the left and Q is on the right.}
\label{residual_map_figure}
\end{center}
\end{figure*}

\subsection{Reconstructed Map Residuals}

The residual map $\epsilon$ between a reconstructed map and the input simulated map can defined for each of the Stokes parameters as

\begin{eqnarray}
\epsilon_{p} &= & x_{p}^{out} - x_{p}^{in} \nonumber\\
&=&\epsilon_{p}^{S} + \epsilon_{p}^{N}
\label{residual equation},
\end{eqnarray}
where $\epsilon_{p}^{S}$ and $\epsilon_{p}^{N}$ represent signal error and pixel noise respectively, $x_{p}^{out}$ is the recovered map estimate and $x_{p}^{in}$ is the input map used to simulate the TOD.  The signal error for the algorithms has been shown to be negligible, so $\epsilon_{p} \rightarrow \epsilon_{p}^{N}$.

\subsubsection{Signal and detector noise simulations}

\begin{figure}
\begin{flushleft}
\includegraphics[angle=0,width=0.49 \textwidth]{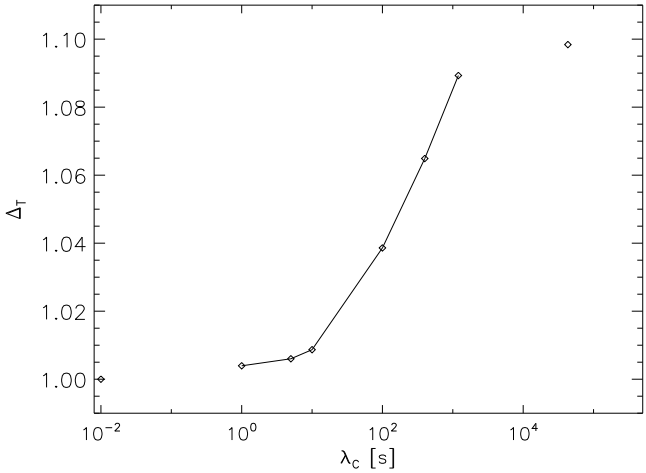}%{plots/delt_ata_0.1.eps}
\caption{Variation of the mean dimensionless $\Delta_{T} = (\sigma / \sigma_{MLE})_{T}$ for 200 detector $f_{knee}= 0.1$-Hz simulations with destriping function length $\lambda_{C}$ in seconds. The diamond at $\lambda_{C}= 0.01$-s is the maximum likelihood solution and the one at $\lambda_{C}= 43200$-s is the naive solution of equation (\ref{naive estimator}).}
\label{delt_ata_0.1}
\end{flushleft}
\end{figure}

We first look at the detector noise simulations with $f_{knee}=0.1$.  Differences between the maps from different algorithms are not perceptible by eye, as the map is dominated by the signal.  However, differences are perceptible in the residual maps.  The top row of Figure \ref{residual_map_figure} contains residual T and Q maps for  Descart (with $\lambda_{C}= 1$-s), whilst the middle row show residuals from the same data mapped by the naive algorithm (in which the TOD is mapped using equation (\ref{naive estimator}) with no pre-filtering). The characteristic $1/f$ noise stripes are visible in the naive T residuals and are notably absent in the Descart T residuals.  The reduction in the correlated noise is best illustrated by the difference map between the destriped and naive maps (bottom left panel of Figure \ref{residual_map_figure}).  The stripes in this map are the correlated $1/f$ noise in the pixel map that have been \emph{removed} by the destriping process, visibly following the scanning pattern.

The Q and U residuals are very close white noise only due to the modulation of the signals out of the low $f_{knee} $ detector $1/f$ so the destriped and naively mapped residuals (top and middle right panels) are very similar.  Their difference maps show a small change in the magnitude of the noise but no striping structure as for T.  The structure in this difference map can be understood by noting that the $1/f$ noise that seeps through to Q and U maps is itself modulated by sinusoidal functions during the de-modulation process.

The maximum likelihood solution produces maps with the smallest possible residuals.  The relevant statistic to measure the ability of the destriper to return maps cleaned of correlated $1/f$ noise is the ratio of the destriped map's root-mean-squared (rms) residual to that of the maximum likelihood map's rms residual, $\Delta{i}$:

\begin{equation}
\Delta_{i}= \Bigg(\frac{ \sigma} { \sigma_{MLE} }\Bigg)_{i}
\label{del definition},
\end{equation}
where $i$ is one of the Stokes parameters T, Q or U.  This dimensionless statistic is independent of the white noise level in the map, it is the constant by which the noise in the map is multiplied above optimality.

Figure \ref{delt_ata_0.1} shows the variation in mean (over 200 realisations) $\Delta_{T}$ with chunk length $\lambda_{C}$ in the $f_{knee}=0.1$-Hz simulations.  The unconnected diamonds are mean $\Delta_{T}$ for the maximum likelihood (corresponding to $\lambda_{C}= 0.01$-s) and naive (corresponding to $\lambda_{C}= t_{int}$, the total integration time of the experiment) maps.  As $\lambda_{C}$ is decreased, $\Delta_{T}$ decays towards optimality at $\Delta_{T}=1$.  The best destriped maps are returned at the smallest chunks size considered ($\lambda_{C}=1$-s) and have only $ 0.4\%$ higher pixel residuals than the optimal map, achieving $96\%$ of the reduction in pixel residuals that the maximum likelihood method brings over naive binning.  These numbers are relative to optimality and are dependent only on the correlated noise.  They are independent of the magnitude of the white noise floor.

The destriping method models the noise as a white component + a correlated component described by a series of offset functions.  The noise covariance $C_{N}$ is modelled as 

\begin{equation}
C_{N} \approx \sigma_{W}^{2} \delta_{tt'} + F C_{a} F^{T}
\label{noise approx}
\end{equation}
where $C_{a}$ is solely responsible for the off-diagonal part of $C_{N}$.  When the resolution of $C_{a}$ is increased, the approximation to the real $C_{N}$ is more accurate and so the residuals are smaller.  The residual reduction flattens as $\lambda_{C}$ is decreased significantly below the noise correlation length (10 seconds for $f_{knee}=0.1$-Hz).

\begin{figure}
\begin{flushleft}
\includegraphics[angle=0,width=0.5 \textwidth]{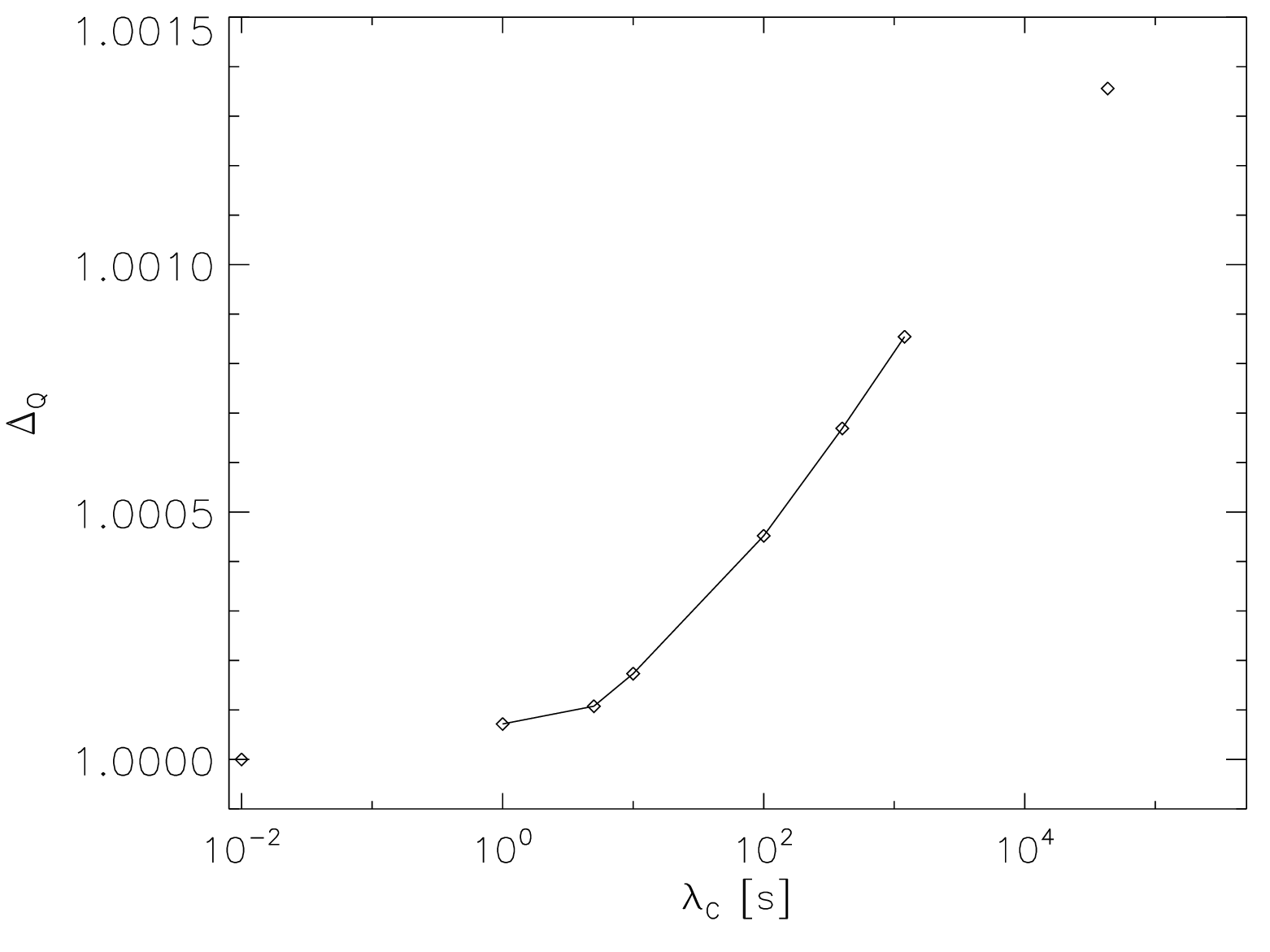}%{plots/delq_ata_0.1.eps}
\caption{Variation of mean dimensionless $\Delta_{Q}$ with destriping baseline length $\lambda_{C}$ in seconds, for the detector simulations.}
\label{delq_ata_0.1}
\end{flushleft}
\end{figure}

Mean $\Delta{Q}$ for these simulations is shown in Figure \ref{delq_ata_0.1}.  Despite the modulation of the Q and U signals to $f_{mod}=20$-Hz (200 times higher than the noise $f_{knee}$), there remains a gradient in the relative noise residuals between the optimal map and the naive map.  Destriping converges to near optimal noise levels at $\lambda_{C}= 1$-s, as for the T maps.  The difference in pixel error between the algorithms is at most of order 0.1\%.

\begin{figure}
\begin{center}
\includegraphics[angle=0,width=0.48 \textwidth]{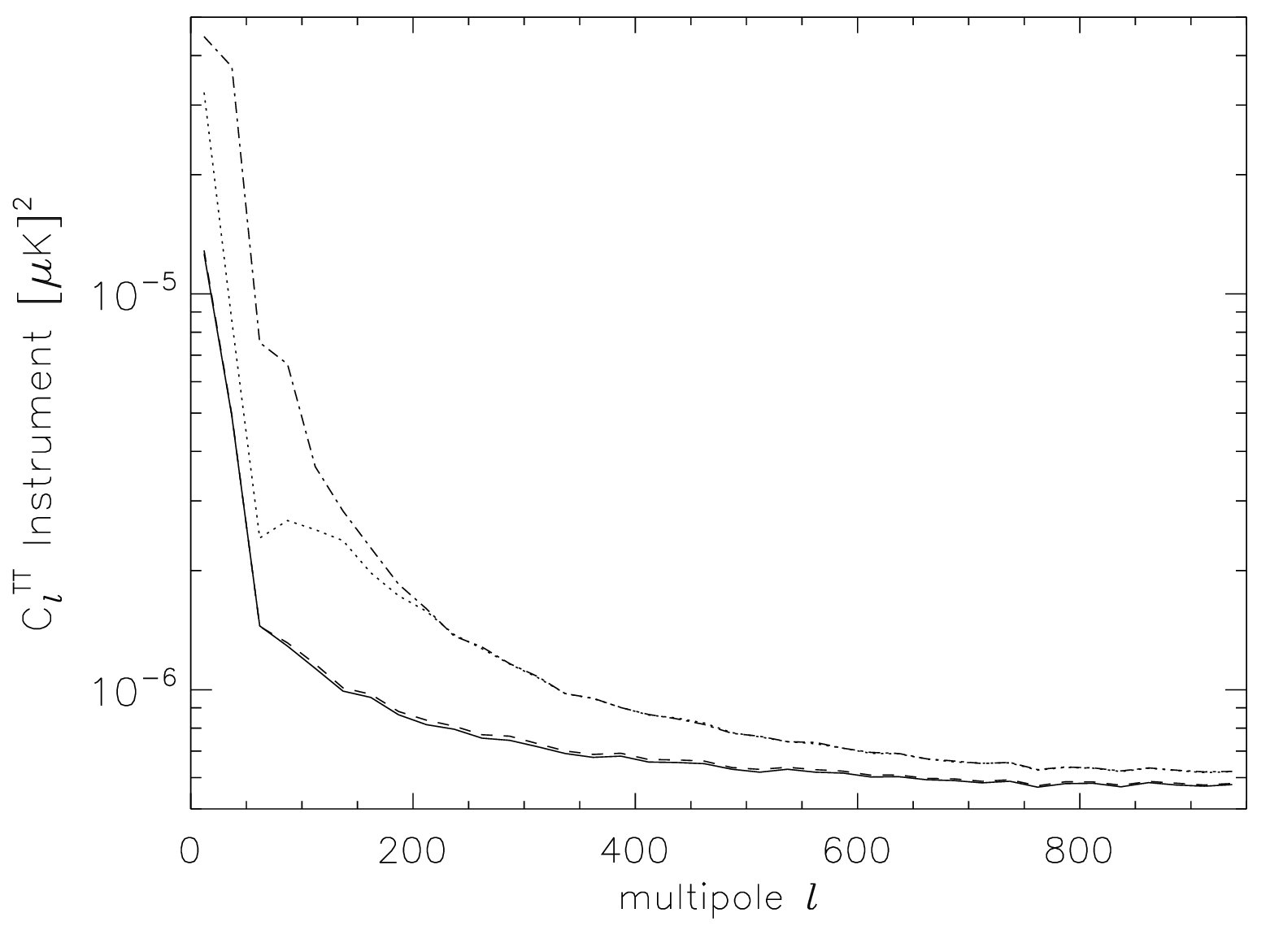}
\includegraphics[angle=0,width=0.49 \textwidth]{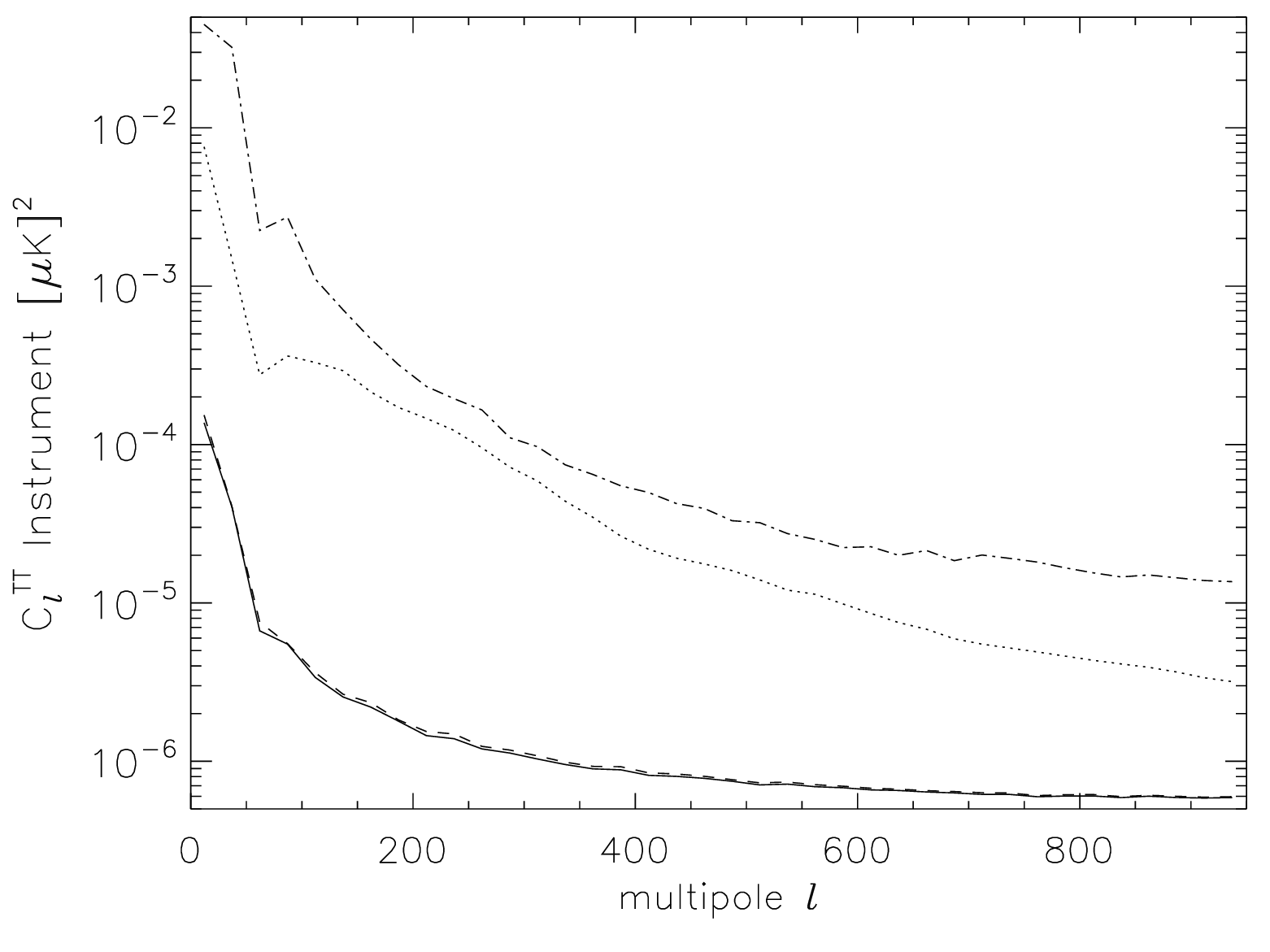}
\caption{\emph{Upper panel}: Mean T residual angular power spectra for the detector $f_{knee}=0.1$-Hz simulations. The solid curve is for the MLE residuals; the dashed curve is for the 1 second destriping residuals; the dotted curve is for the 400 seconds destriping residuals; the dot-dashed curve is for the naive binning (equation \ref{naive estimator}) residuals. \emph{Lower panel}:  Same as above, but for the mean T residual angular power spectra of the signal+atmospheric noise simulations.}
\label{mean_t_clres}
\end{center}
\end{figure}

The extra noise power in the naive T maps is at all angular scales, as shown in the residual angular power spectra in the upper panel of Figure \ref{mean_t_clres}.  The curves in this plot are the mean T residual angular power spectra for all 200 realisations.  The residual power spectrum for the optimal maximum-likelihood algorithm is the solid curve.  The long dashed curve is from the Descart 1-s residuals, the dotted curve from the Descart 400-s residuals and the dot-dashed curve is from the naive residuals.  The 400-s baseline Descart spectrum is included as an example of ``quick and dirty" destriping as opposed to using destriping to replace the maximum-likelihood approach.  This coarser noise model makes some improvement to noise at large scales but fails to reduce noise power at scales beneath its $\mathbf{C_{a}}$ resolution.  The best 1-s baseline Descart maps have residuals nearly indistinguishable from optimality at all angular scales.

\subsubsection{Signal and atmospheric noise simulations}

\begin{figure}
\begin{flushleft}
\includegraphics[angle=0,width=0.49 \textwidth]{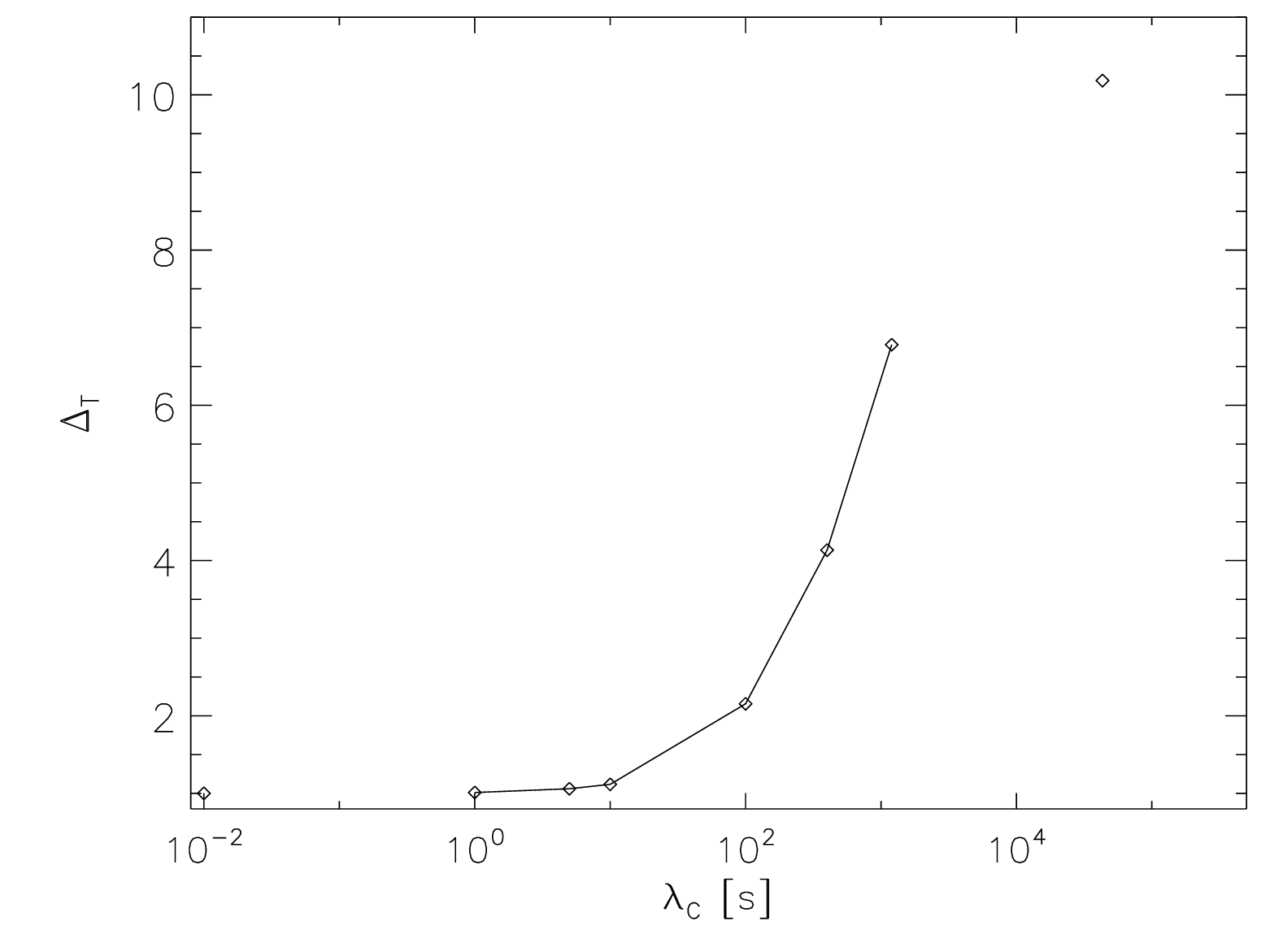} \\%{plots/delt_ata_0.1.eps}
\includegraphics[angle=0,width=0.49 \textwidth]{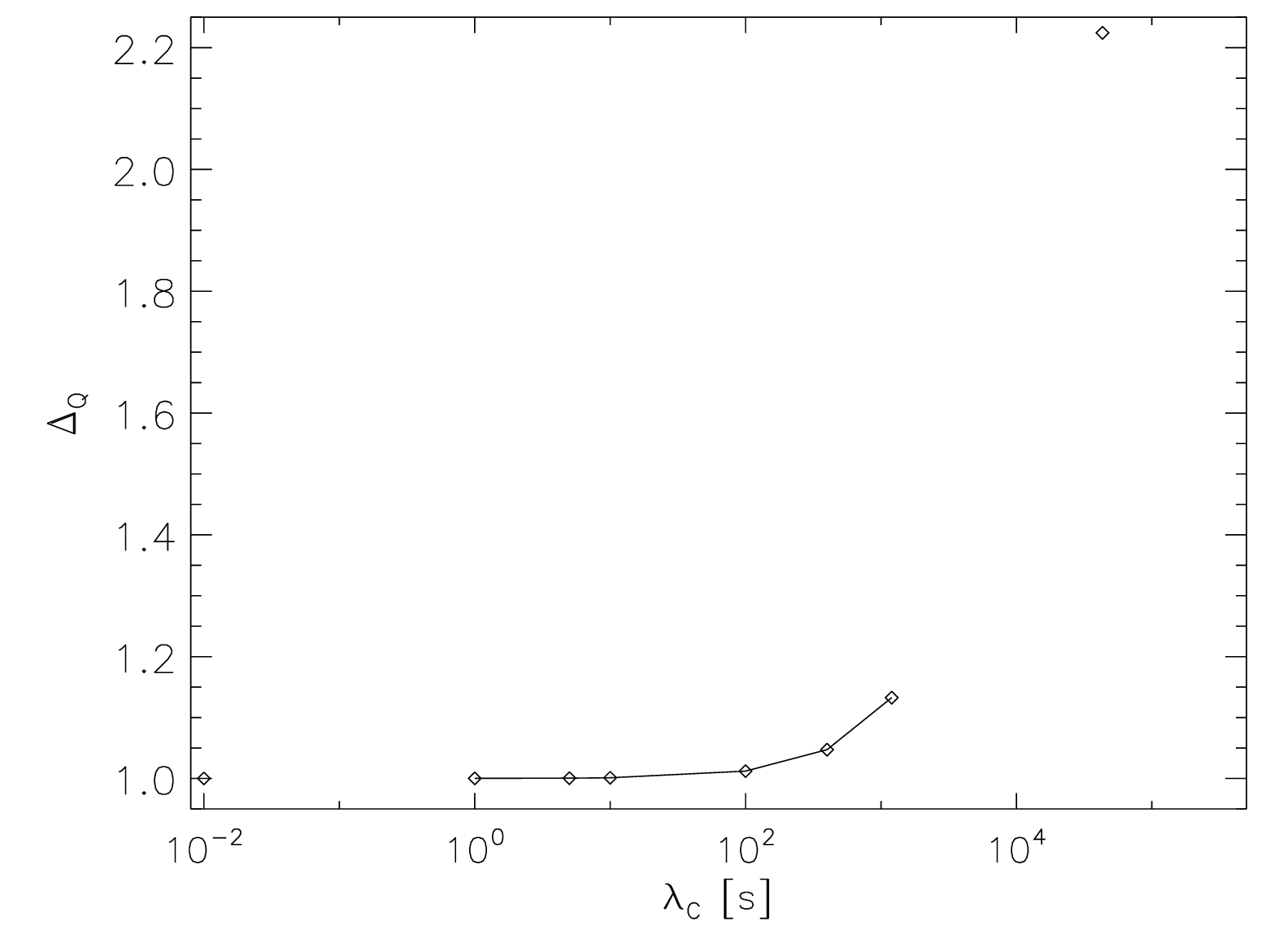}
\caption{\emph{Upper panel}: Variation of mean dimensionless $\Delta_{T}$ for the signal + atmospheric noise simulations. \emph{Lower panel}: Variation of mean dimensionless $\Delta_{Q}$ for the signal + atmospheric noise simulations.}
\label{delt_ata_atmos}
\end{flushleft}
\end{figure}

The simulated atmospheric noise has more correlated noise, with $1/f$ $f_{knee}=0.2$-Hz and a much larger spectral index of $\alpha=1.9$.  Correspondingly, significantly more correlated noise leaks through to the T map from naive map-making.  In this case, sophisticated mappers are essential, reducing the rms T residual by a factor of 10 and the Q residuals by a factor of 2 (Figure \ref{delt_ata_atmos}).  The majority of the improvement is achievable using a baseline of 10-s, achieving $98.7\%$ of the possible improvement in T residual power and $99.0\%$ of the possible improvement in Q residual power.  The best 1-s baseline maps achieve $99.9\%$ of the possible improvement for T and $99.9\%$ of the possible improvement for Q.

Despite the hardware effort of QU modulation using a half-wave plate, this noise regime produces reducible Q and U pixel noise sourced from correlated TOD noise that can be more than halved by the application of even relatively long baseline destriping.  This assumes no $T \rightarrow P$ mixing from instrumental polarisation, the results of which would alias stripes into the Q and U maps themselves, strengthening the requirement for destriping.

The lower panel of Figure \ref{mean_t_clres} shows the mean angular power spectra of the T residuals for the algorithms over 200 realisations (the curves are the same as those in the upper panel of Figure \ref{mean_t_clres}).  It should be noted that the optimal map shows slightly more noise power for these simulations than it did in the detector simulations, because even the optimal maximum-likelihood solution cannot remove all of the $1/f$ noise.

The ``quick and dirty" 400-s baseline approach is more effective at smaller angular scales than it was for the detector noise.  The best 1-s baseline Descart maps are again nearly indistinguishable from optimal maps at all angular scales.

\subsubsection{Fence scan simulations}

\begin{figure}
\begin{flushleft}
\includegraphics[angle=0,width=0.49 \textwidth]{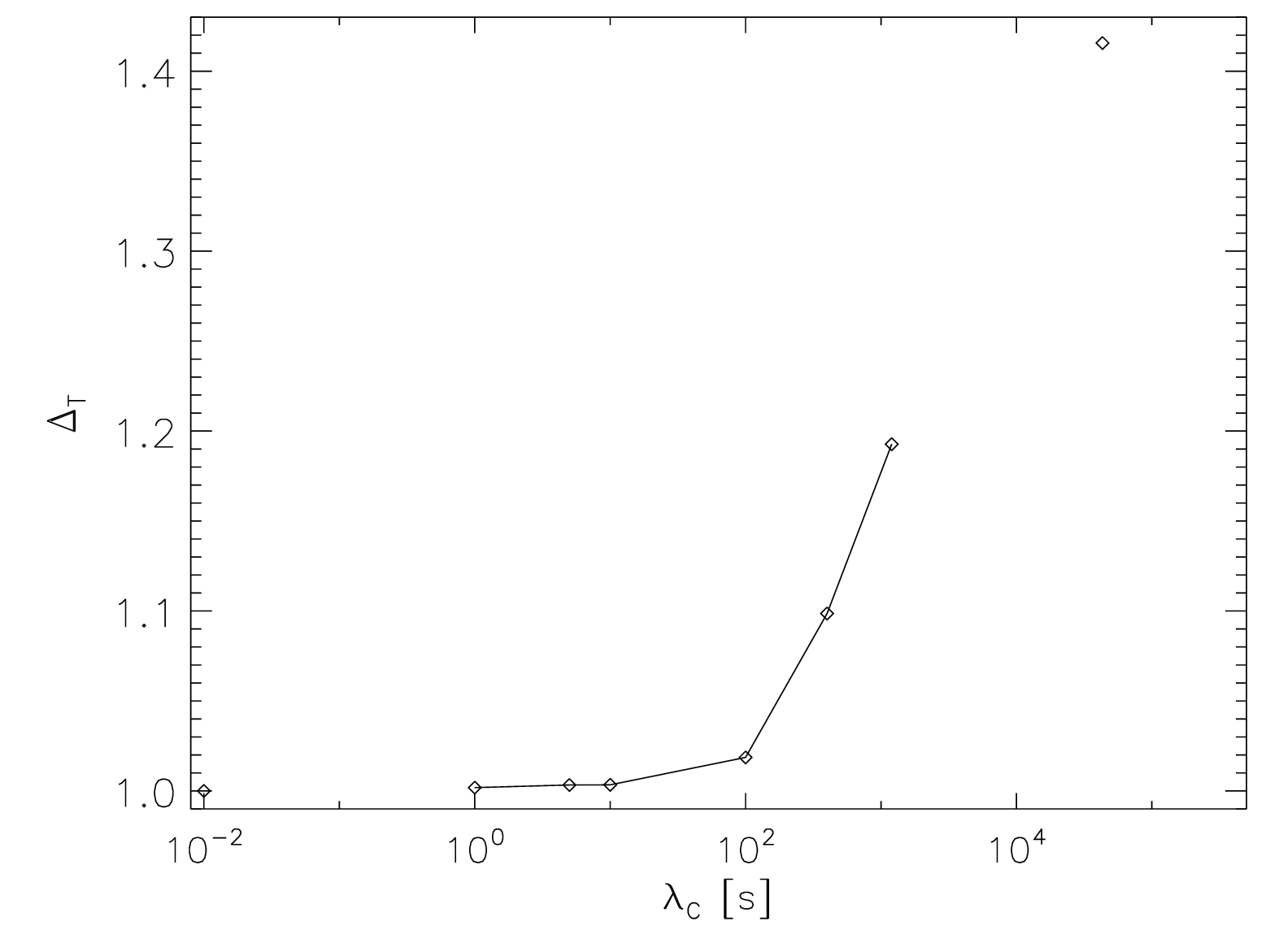}
\includegraphics[angle=0,width=0.49 \textwidth]{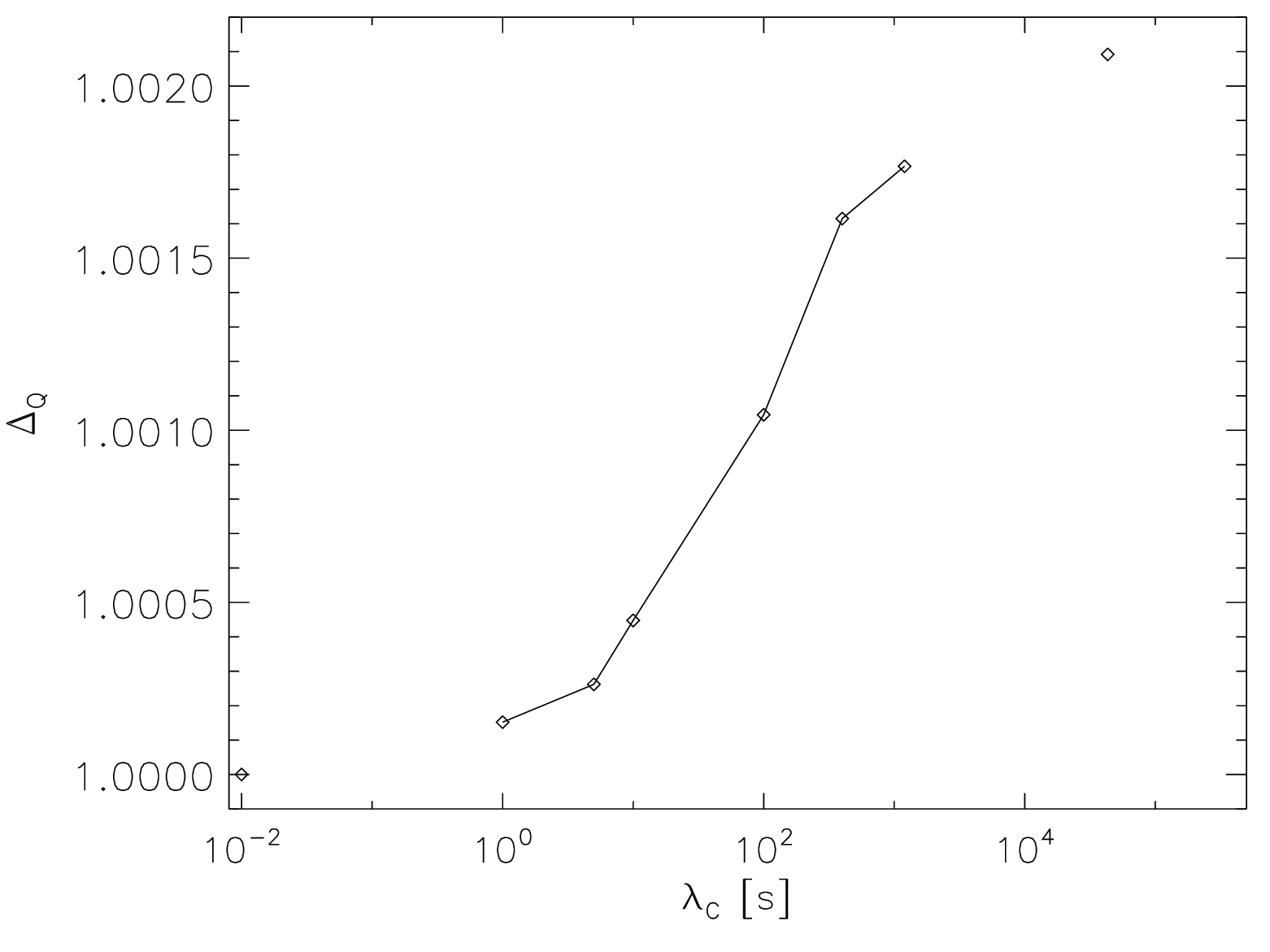}
\caption{\emph{Upper panel}: Variation of dimensionless $\delta_{T}$ for the fence scan simulations. \emph{Lower panel}: Variation of dimensionless $\delta_{Q}$ for the fence scan simulations. }
\label{fence rms residuals}
\end{flushleft}
\end{figure}

The fence scan simulations used the same noise parameters as the signal and detector noise simulations with $f_{knee}=0.1$-Hz and $\alpha=1.0$.  The scan was designed to cover approximately the same number of pixels using the same number of observations.  The average integration time per pixel is the same, however the isotropy and connectivity of the scans are different.  Both the fence and sabre scans are sinusoidal scans, spending disproportionally more time integrating at the edges of the field, the fence scan integrating more at all four field edges and the sabre scan integrating more only on two edges.  The connectivity, or level of cross-linking, is much greater in the fence scan, with the two scanning directions ideally cross-linked in being perpendicular to one another.

The rms residuals in the optimal fence maps are $8.7\%$ smaller than the optimal sabre maps.  Further to this, Figure \ref{fence rms residuals} shows that the Descart T residuals decay toward optimality at longer baselines than for the sabre scan, reaching a trough at 10-s baselines: further reductions yield negligibly different residuals as the maps are already near-optimal.  

The improvements in Q residuals are similar to the improvements in the sabre scan simulations: the level of cross-linking has little effect on the near-white noise in the modulated Q and U streams.

\subsection{E and B mode errors}

\begin{figure}
\begin{center}
\includegraphics[angle=0,width=0.48 \textwidth]{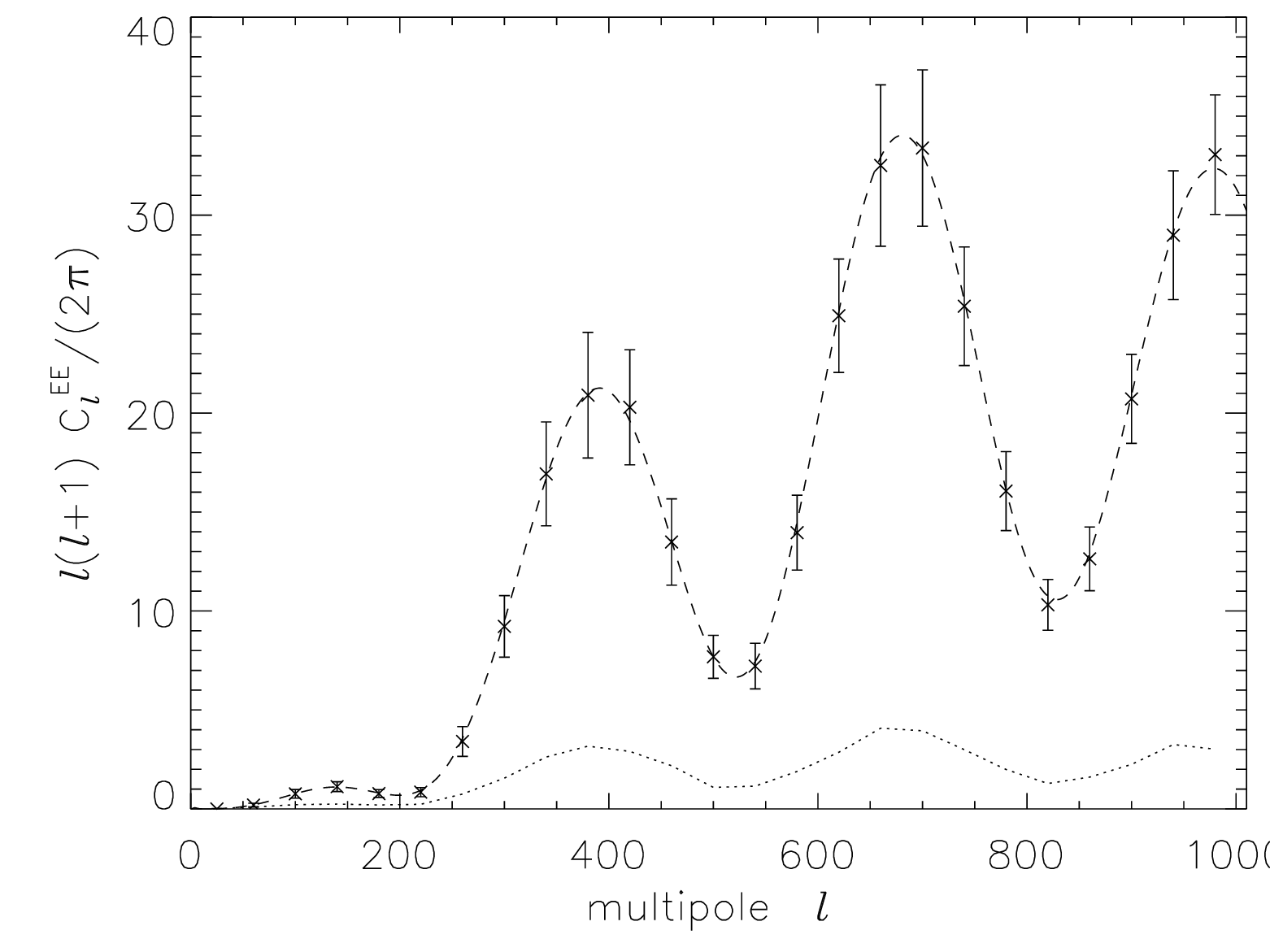}%{plots/delt_ata_0.1.eps}
\caption{Plot of the mean estimated $C_{l}^{EE}$ for the $f_{knee}=0.1$-Hz simulations, using Descart on 200 noise and 200 signal + noise realisations with 1-s baselines.  The error bars are the average for each individual signal+noise realisation.  The dashed curve is the input power spectrum and the dotted curve is the magnitude of the plotted error bars.}
\label{e plot}
\end{center}
\end{figure}

\begin{figure}
\begin{center}
\includegraphics[angle=0,width=0.48 \textwidth]{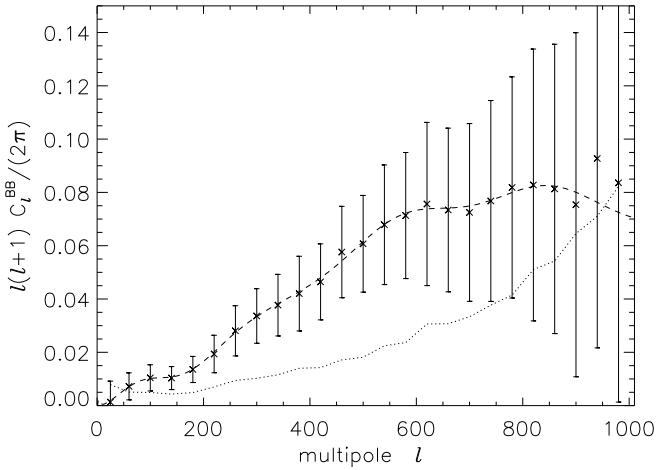}
\caption{As Figure \ref{e plot} but for $C_{l}^{BB}$estimates from Descart 1-s maps.}
\label{b plot}
\end{center}
\end{figure}

The polarisation modulation design is intended to remove the effects of correlated noise from the Q and U maps and thence from the E and B power spectra. However, the increased variance in the Q and U maps from simple naive binning of a single detector propagates through to the polarisation power spectra.

Figures \ref{e plot} and \ref{b plot} show the mean estimated E and B-mode power spectra from the 200 signal+detector noise simulations respectively, after subtraction of the mean noise power spectrum.  The error bars are the Monte-Carlo error bars for a single signal+noise realisation, and the dashed curves are the input theoretical power spectra.  The dotted curve indicates the magnitude of the plotted error bar in each bin, which include both experimental noise and sample variance from partial sky coverage. The mean estimates are unbiased with respect to the input model.

\begin{figure}
\begin{center}
\includegraphics[angle=0,width=0.48 \textwidth]{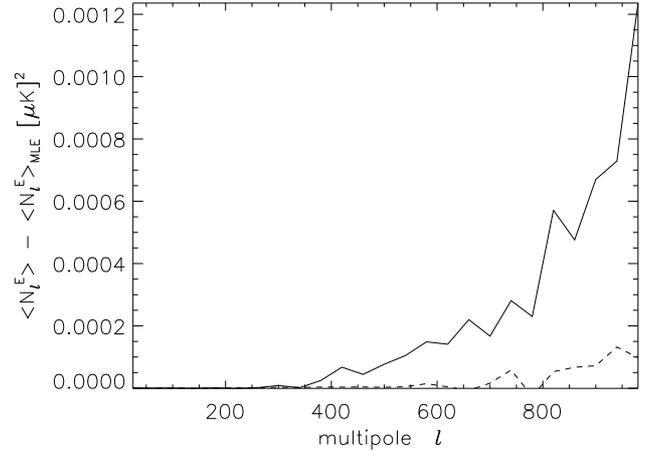}
\includegraphics[angle=0,width=0.48 \textwidth]{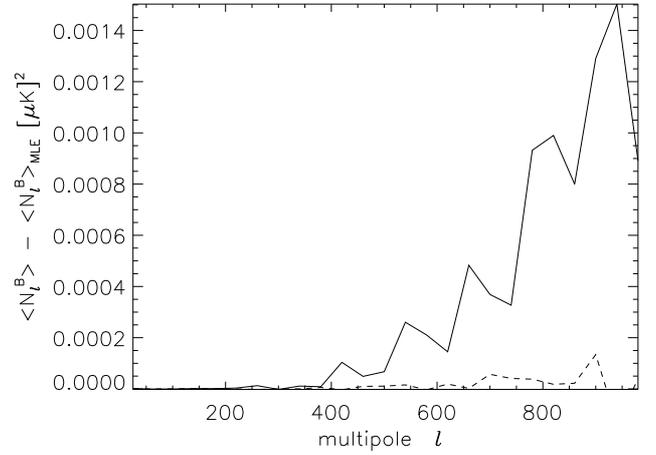}
\caption{\emph{Upper panel}: Differences between E-mode noise power spectra from the $f_{knee}=0.1$-Hz simulations in $\mu K^{2}$.  The solid curve is $\langle N_{l}^{E} \rangle _{naive} - \langle N_{l}^{E} \rangle_{MLE}$ whilst the dashed curve is $\langle N_{l}^{E} \rangle _{destriped} - \langle N_{l}^{E} \rangle_{MLE}$ for Descart with 1-s baselines. \emph{Lower panel}: Differences between B-mode noise power spectra from the $f_{knee}=0.1$-Hz simulations in $\mu K^{2}$.  The solid curve is $\langle N_{l}^{B} \rangle _{naive} - \langle N_{l}^{B} \rangle_{MLE}$ whilst the dashed curve is $\langle N_{l}^{B} \rangle _{destriped} - \langle N_{l}^{B} \rangle_{MLE}$ for Descart with 1-s baselines.}
\label{fk0.1 ebnoise}
\end{center}
\end{figure}

The mean noise angular power spectra from the signal+detector noise simulations are shown in Figure \ref{fk0.1 ebnoise}.  The curves are the difference between the noise angular power spectra either from naive (solid curve) or destriping (dashed curve) and that of the maximum-likelihood noise power spectrum, defined by: $\langle N_{l} \rangle^{approx} - \langle N_{l} \rangle^{MLE} $, where $\langle N_{l}\rangle^{approx}$ is the noise spectrum from either the 1-s Descart maps or the naive maps.  If either is optimal, the curve will be all zeros.

Whilst neither curve is optimal, the change in noise power is orders of magnitude smaller than the signal power at all multipoles and has a negligible effect on the error bar magnitudes. 

\begin{figure}
\begin{center}
\includegraphics[angle=0,width=0.48 \textwidth]{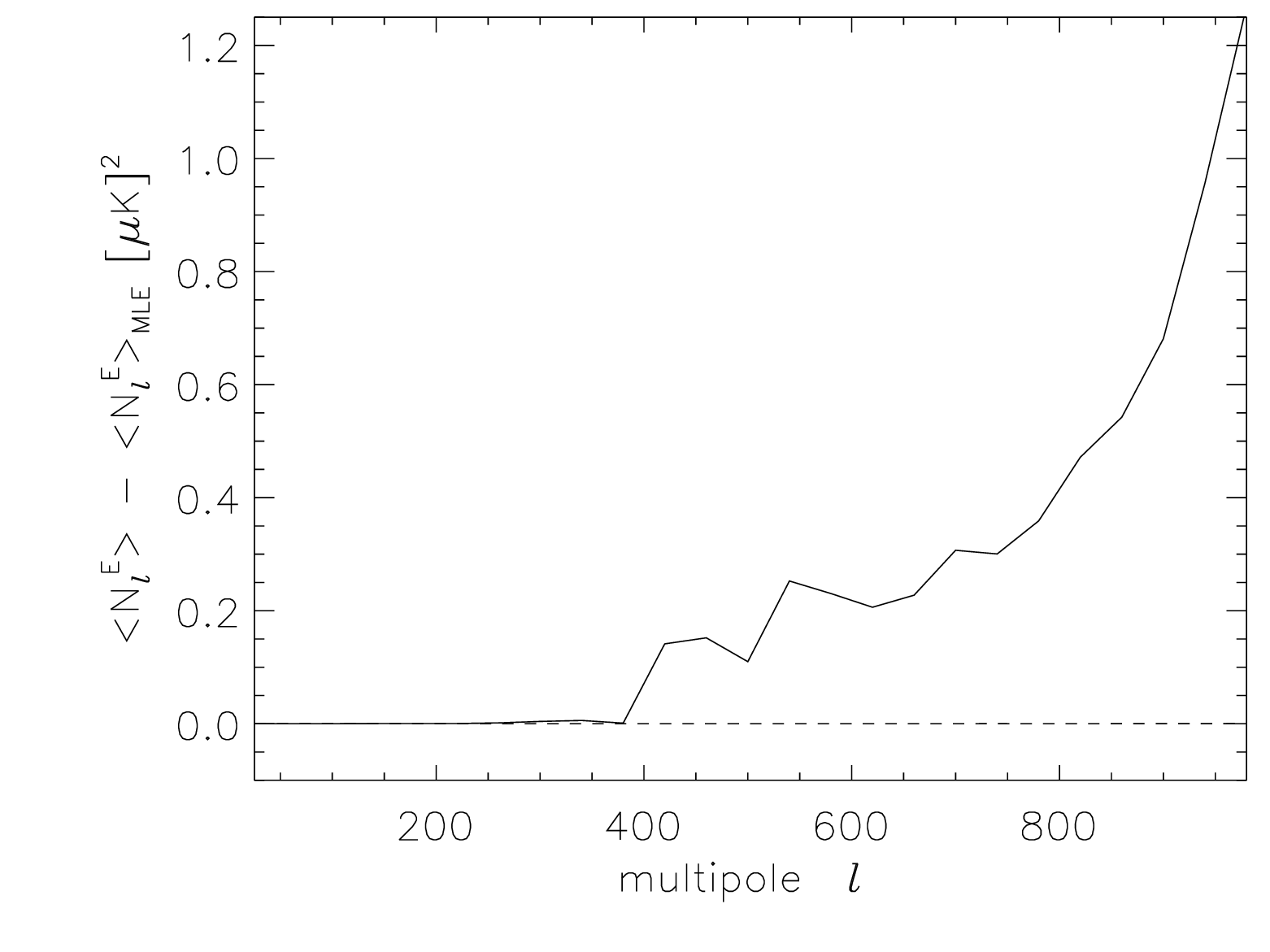}
\includegraphics[angle=0,width=0.48 \textwidth]{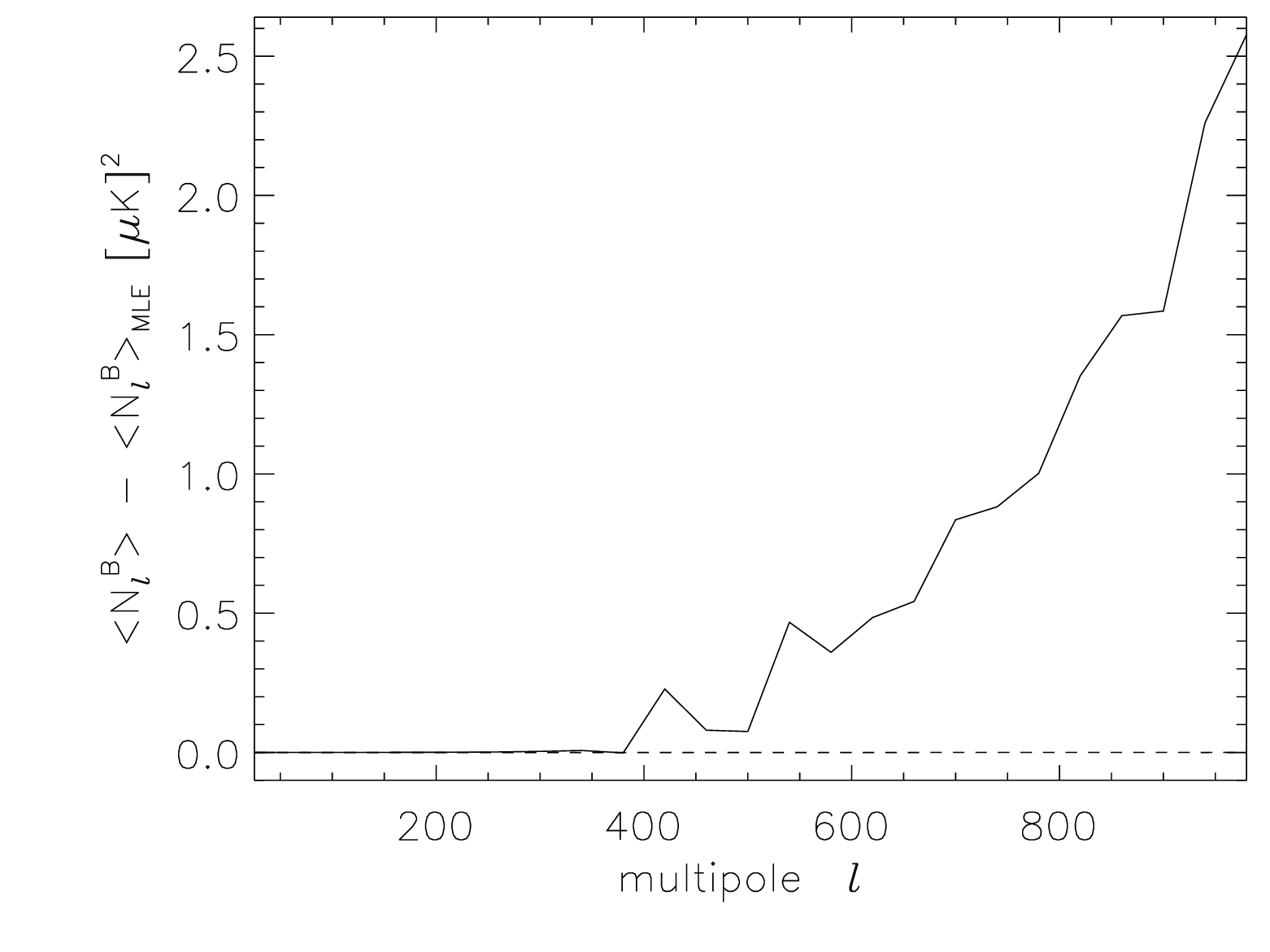}
\caption{\emph{Upper panel}: Same as the upper panel of Figure \ref{fk0.1 ebnoise}, but for the atmospheric noise simulations. \emph{Lower panel}: Same as the lower panel of Figure \ref{fk0.1 ebnoise}, but for the atmospheric noise simulations. }
\label{fk_atmos ebnoise}
\end{center}
\end{figure}

Figure \ref{fk_atmos ebnoise} shows the noise angular power spectra difference for the signal+atmospheric noise simulations, where the changes are very significant.  At multipoles $l > 400$, the B-mode noise angular power spectrum becomes larger than the input signal angular power spectrum.

The effect of this on the B-mode error bars is clearly seen in the significance of the total detection.  The total significance estimator $\hat{C}$ bins the bandpowers of the spectrum ($\hat{C_{b}}$) into a single bin, weighted proportionally to the power of the input fiducial model $C_{b}^{fid}$ and inversely proportional to the variance of the bandpower $\sigma_{b}^{2}$

\begin{equation}
\hat{C} = \frac{\sum_{b}  \frac{C_{b}^{fid}}{\sigma_{b}^{2}}  \hat{C_{b}}}
	 {\sum_{b}  \frac{(C_{b}^{fid})^2}{\sigma_{b}^{2}} }
\label{total sign estimator}.
\end{equation}

The significance of the detection is then given by $\frac{\langle \hat{C} \rangle}{\sqrt{\langle (\hat{C} - \langle \hat{C} \rangle)^{2} \rangle}}$.

The significance of the E and B-mode detections for the algorithms with atmospheric noise are shown in Table \ref{sigma table}.  The increased noise from the naive maps slightly erodes the significance of the E-mode detection and destroys the detection of B-modes.  We re-iterate that this applies only in the case of a single detector and would not apply, for example, to detector differencing experiments.

\begin{table}
\caption{Atmospheric noise simulation ($f_{knee}=0.2, \alpha=1.9$), detection significance}
\begin{center}
\begin{tabular}{c|c|c}
&E-mode&B-mode \\
\hline
MLE & $37.90$ & $8.22$ \\
Descart 1-s &$37.90$&$8.22$ \\
Naive&$35.29$&$0.02$\\
\hline
\end{tabular}
\end{center}
\label{sigma table}
\end{table}%

\subsection{Spurious B modes}

200 signal+noise simulations were created using input CMB maps with artificially zero B-modes and typical detector $1/f$ ($f_{knee}=0.1$-Hz).  Maps were made from these TOD streams using each algorithm and the B-modes of the polarisation fields were estimated as in the previous section.

\begin{figure}
\begin{center}
\includegraphics[angle=0,width=0.48 \textwidth]{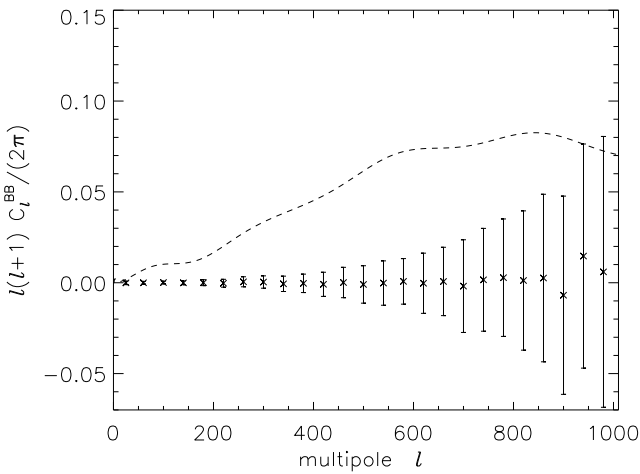}
\caption{As Figure \ref{b plot} but for the simulation with zero $C_{l}^{BB}$ from Descart 1-s maps.}
\label{nob plot}
\end{center}
\end{figure}

There was no evidence of the presence of spurious B-modes due to mode mixing from any of the algorithms.  Figure \ref{nob plot} shows the mean estimated B mode for these TOD for Descart with 1-s baselines (the plots are identical for the naive mapper and the MLE algorithm). It is unbiased, correctly returning an ensemble average zero B mode estimate.  

\subsection{Computing resources}

\begin{figure}
\begin{center}
\includegraphics[angle=0,width=0.45 \textwidth]{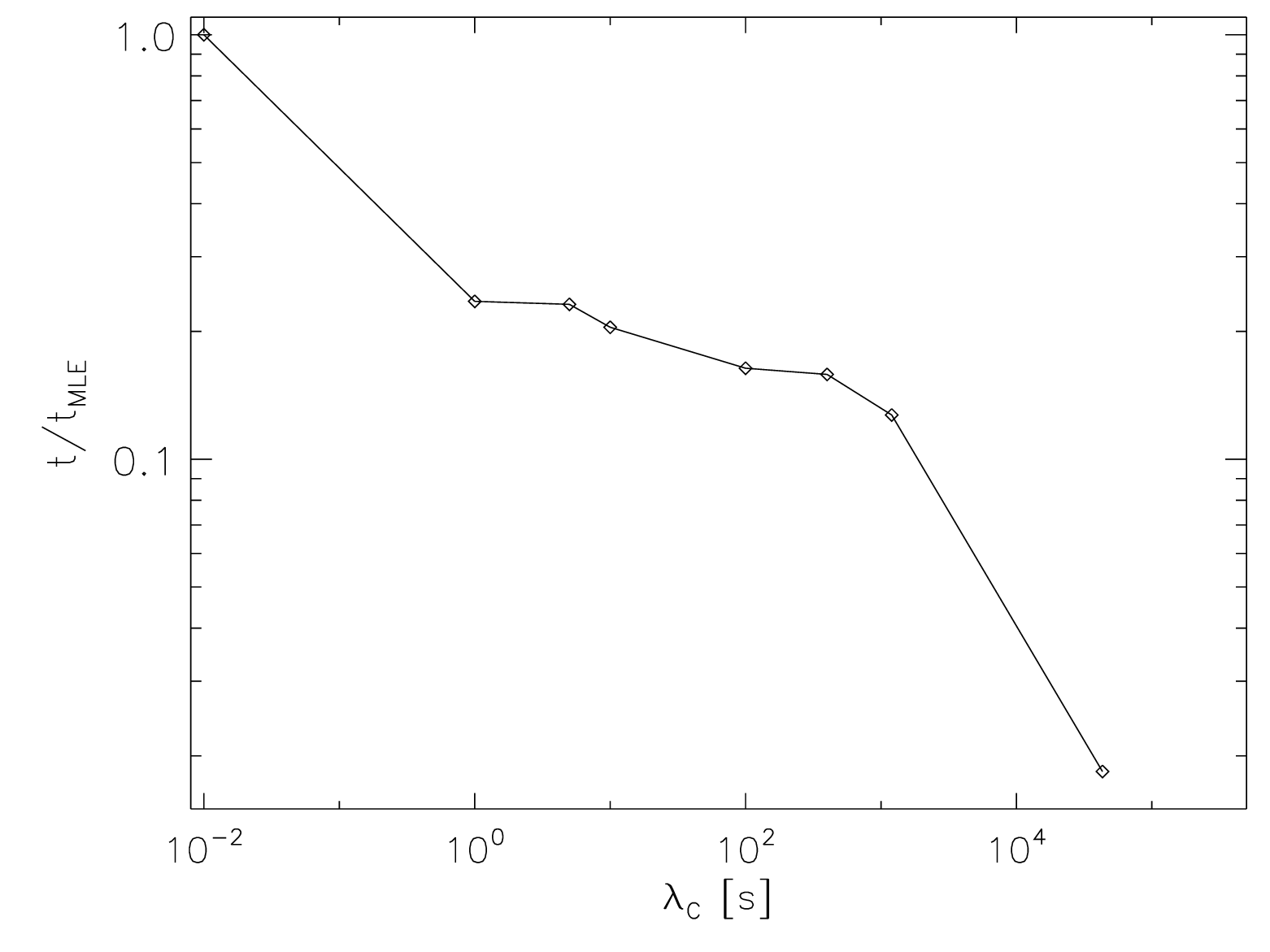} \\%{plots/runtime_ata_0.1.eps}
\includegraphics[angle=0,width=0.45 \textwidth]{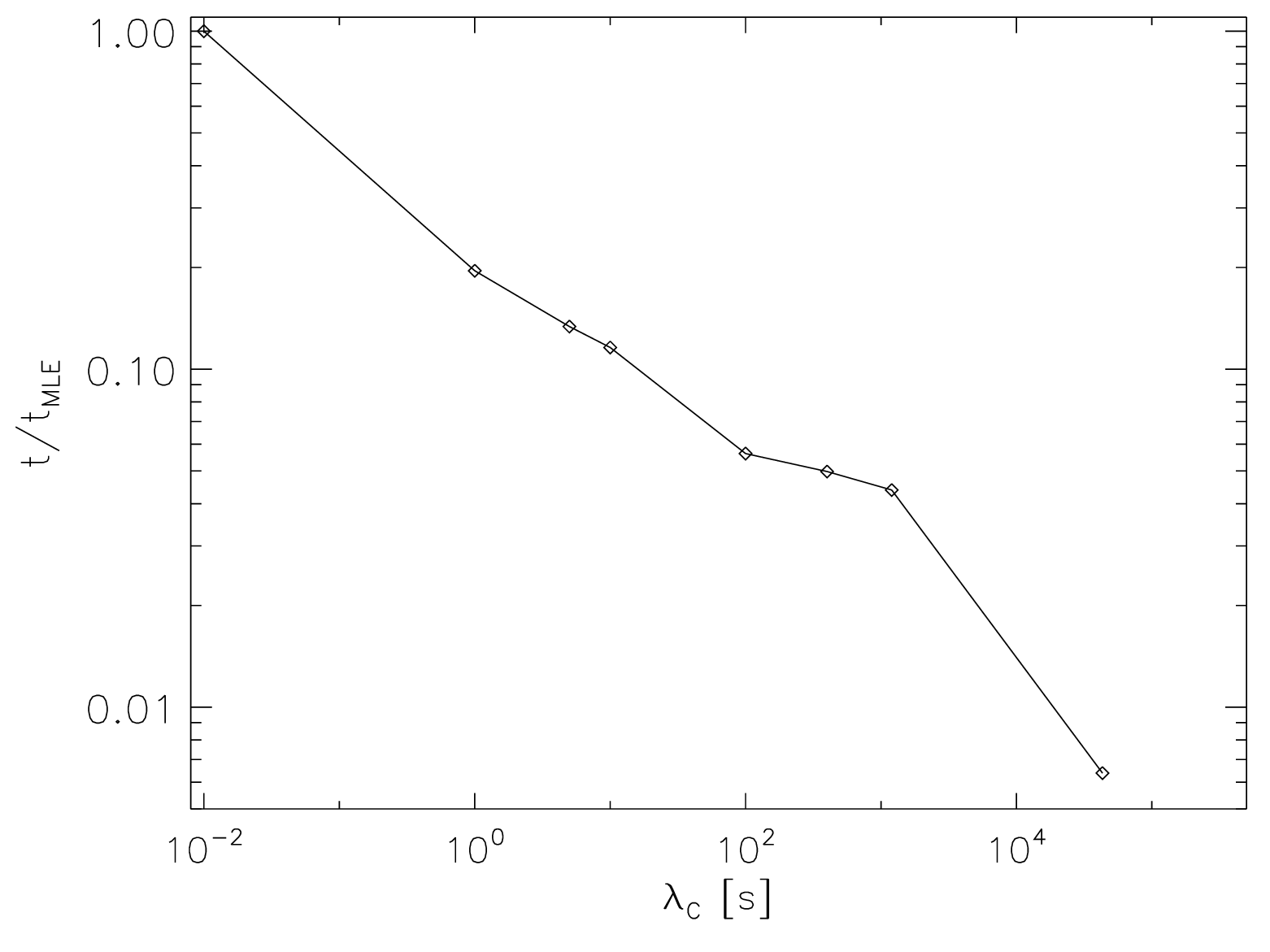}
\caption{Comparative runtime ($t/t_{MLE}$) variation with destriping length $\lambda_{C}$.  \emph{Upper panel}: runtimes for the signal+detector noise simulations. \emph{Lower panel}: runtimes for the signal+atmospheric noise simulations.}
\label{runtime_ata_0.1}
\end{center}
\end{figure}

The principal concern for computing resources for map-making is CPU time, as it feasible to hold many single days' worth of TOD in memory at one time.  The naive algorithm represents the quickest, and dirtiest, possible method for reducing TOD into a map.  Its computation time is dominated by the input/output overhead.

The iterative scaling of the destriping algorithm is $\mathcal{O}(N_{t} + n_{a} \log_{2}|n_{a}|)$,compared to $\mathcal{O}(N_{t} (1 + \log_{2}|N_{t}|))$, for the maximum likelihood algorithm (where $N_{t}$ is the number of number of observations).  Effective values of $n_{a}$ always satisfy $n_{a} << N_{t}$, for example, the $n_{a}$ used in this paper are between 2 and 5 orders of magnitude smaller than $N_{t}$, so that the iterative scaling of Descart $N_{t} +n_{a} \log_{2} |n_{a}| \approx N_{t}$.  Noting that typical time stream lengths for day strategies are of order $10^{7}$, we can predict that destriping will require approximately an order of magnitude less CPU time to run.

In practice, the time-stream can be split into chunks to speed up the Fourier de-convolutions, where the chunk size is some multiple of the correlation length of the time-stream noise. For a chunk size of $n_{c}$ the iterative scaling of the FFTs reduces to $\mathcal{O}(N_{tod}\log_{2} |n_{c}|)$ for the maximum-likelihood algorithm and $\mathcal{O}(N_{t} +n_{a} \log_{2} |n_{ca}|)$ for destriping algorithm, where $n_{ca} = n_{c} / \lambda_{c}$ is still $2-5$ orders of magnitude smaller than $n_{c}$, maintaining the reduced complexity of the destriping algorithm.  We also note that the use of hardware optimised  FFT libraries, as opposed to the ubiquitous FFTW library, will further accelerate the FFT procedures.

Scaled computing times ($t/t_{MLE}$, where $t_{MLE}$ is the CPU time required for the maximum likelihood algorithm to complete) for the algorithms are shown in Figure \ref{runtime_ata_0.1} for both the signal+detector noise simulations and the signal+atmospheric noise simulations.  The codes are all serial and the runs were completed on the same machine, using dual-core Intel Xeon 3GHz CPUs, with the same allocated resources.  The longest Descart baseline used was 1200-s, which returned a $49.1\%$ reduction in Q and U residuals for the atmospheric noise simulations (Figure \ref{delt_ata_atmos}, lower panel), out of a possible $55.04\%$ for the optimal algorithm, whilst achieving a $22\times$ speed improvement over the optimal algorithm.

\begin{table}
\caption{Typical numbers of Descart iterations for the signal+detector noise, signal+atmospheric noise and fence scan simulations.  Also shown is the destriping baseline length $\lambda_{C}$ in seconds and the corresponding number of offset functions in the system ($n_{a}$).}
\begin{center}
\begin{tabular}{c|c|c|c|c}
$\lambda_{C}$ (s)& $n_{a}$ &detector noise &atmospheric noise &fence scan \\
\hline
1     & 43200 &40&127& 18\\
5     & 8640&36& 89& 15\\
10   &4320&32& 65& 15\\
100 & 432 & 22&25& 11 \\
400 & 108 &16&19& 9 \\
1200& 36 &10&11& 7 \\
\hline
\end{tabular}
\end{center}
\label{iterations table}
\end{table}%

In addition to the iterative scaling, the computing time is also determined by the number of iterations required to solve the system.  Table \ref{iterations table} shows the typical numbers of iterations required by Descart for the simulations compared to the destriping baseline length used.  The fence scan simulations required fewer iterations than the sabre scan simulations with the same noise parameters (detector noise column).  The greater cross-linking of the fence scan has produced a system that is easier to solve than the sabre scan especially at short baseline lengths where the system ($n_{a}$) becomes large.

The number of iterations is similar for both sets of sabre scan simulations (detector noise and atmospheric noise columns) at very long baselines, but for the atmospheric noise increases much more quickly with baseline lengths $<100$-s, requiring $\approx 3 \times$ as many iterations than the detector noise simulations at the near optimal 1-s baseline.

By modelling the correlated noise as a series of offset functions, the destriping algorithm is solving a smaller system than the full maximum likelihood algorithm (generally $n_{a} < n_{p}$ except for the very shortest baselines, where $n_{a}$ and $n_{p}$ are number of offsets and pixels respectively).  Typical experimental resolutions in future experiments will be higher than that used here ($n_{side}=512$), for example $n_{side}= 2048$ for C$_{l}$OVER, increasing $n_{p}$ by a factor of 16 and leaving $n_{a}$ untouched.

\subsection{Importance of including $C_{a}$}
\label{Ca section}

\begin{figure}
\begin{flushleft}
\includegraphics[angle=0,width=0.49 \textwidth]{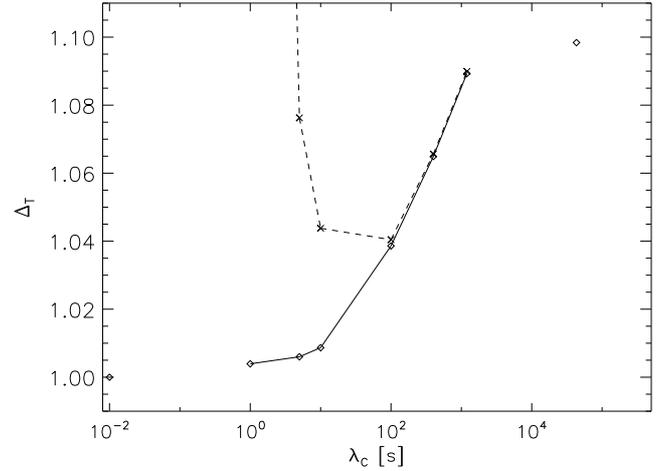}%tra_cov_plot.eps}
\caption{Comparison between destriping with (solid line with circles) and without (dashed line with $\times$) $C_{a}$ information.}
\label{tra_cov_plot}
\end{flushleft}
\end{figure}

Prior to the development of the MADAM algorithm \citep{keihanen:2005}, destriping was conducted without prior noise information.  This ``traditional" destriping used only the \emph{data term} in (\ref{simple destriping chisq}) (\cite{burigana:1997}, \cite{delabrouille:1998}, \cite{maino:1999} \cite{keihanen:2004}), negelecting the \emph{prior term} containing $C_{a}$.  Without this noise information, the potential of destriping to remove correlated noise is considerably reduced.  It is not possible to use traditional destriping at $\lambda_{C}$s where there is significant correlations between the offset function amplitudes $a_{i}$.

Figure \ref{tra_cov_plot} shows the effects of using traditional destriping in this regime.  The dashed line shows the returned destriping $\Delta_{T}$ vs $\lambda_{C}$.  At large $\lambda_{C}$, where the offsets are uncorrelated, traditional destriping performs similarly to the covariant destriping using $C_{a}$ (the solid curve with circles).  When $\lambda_{C}$ is reduced into the regime of correlated offsets, the residuals behave pathologically, exploding to considerably higher magnitude than for the naive maps.

This can be explained in two ways.  The noise model in the data term of (\ref{simple destriping chisq}) assumes no correlations between the offsets.  When correlations are present, the noise model breaks down and the results are junk.  Including $C_{a}$ introduces prior information on noise correlations into the model.  This can also be recognised as linear regularisation, where the introduction of the regularising matrix $C_{a}$ forces the solution to the system to be well behaved.

This highlights the vital importance of including $C_{a}$.  With its inclusion, destriping can be used to its full potential in returning near optimal maps at a fraction of the computing time.

\section{Discussion}

In this paper, we set out to compare the destriping approach to map-making to the optimal maximum likelihood estimator.  In particular, we want to know if destriping is fast and accurate enough to replace the maximum likelihood estimator for partial-sky, polarisation modulation experiments.

We have written a new covariant destriping code named Descart (DEStriping CARTographer) and a new maximum-likelihood estimation code and applied them to non-circular scanning strategies for the first time.  We introduce variations in cross-linking through our realistic, but non-ideal, sabre scan and the unrealistic, but ideally cross-linked, fence scan.

Neither of these strategies allow the co-addition of circular scans before destriping, a process relied upon by  the early ``traditional" destriping efforts (\citealt{burigana:1997}, \citealt{delabrouille:1998}, \citealt{maino:1999}, \citealt{keihanen:2004}).  We must use a ``covariant" destriping algorithm \citep{keihanen:2005}, incorporating prior knowledge of the noise power spectrum through $\mathbf{C_{a}}$.  This allows the use of very short baselines that produce nearly optimal residuals in the output T, Q and U maps.  We have shown that, without this prior, the noise model clearly breaks down pathologically at shorter baselines and the returned map becomes junk.

The simulations we produced applied destriping to different noise regimes than those to which it has previously been applied.  We have probed regimes of both detector noise and atmospheric noise, the latter of which represents a much greater challenge as it dominates noise even in the current generation of ground-based polarisation experiments.  Previous work on the efficacy of destriping concentrated on the Planck satellite mission, assuming low levels of correlation in the $1/f$ noise, as is possible in the absence of the atmosphere, using knee frequencies of $0.03$-Hz for the HFI \citep{ashdown:2007a} and $0.05$-Hz for the LFI \citep{ashdown:2007b}.  We have simulated detector noise with $f_{knee}=0.1$-Hz and atmospheric noise with $f_{knee} = 0.2$-Hz  and a spectral index $\alpha=1.9$, similar to the level experienced by QuAD \citep{hinderks:2008}.

Destriping performs better in the more realistic and challenging atmospheric noise simulations than in the detector only noise simulations.  For the shortest baseline of 1-s, it achieves $99.9\%$ of the possible improvement in T, Q and U residuals achievable with the full maximum-likelihood algorithm.  This amounts to reduction in T residual magnitude of $90.02\%$ and in Q and U of $55.04\%$.  One of the motivators of this study is the requirement to produce high quality T maps, in addition to the Q and U maps, in order to remove $T\rightarrow P$ leakage from instrumental polarisation (eg: \citealt{johnson:2007}), which will also leak atmospheric $1/f$ into the Q and U signal despite the modulation of those signal by the rotating half-wave plate.  

We have found that even without the simulation of this leakage, it is vital to deal with the correlated noise in order to observe B-modes: with naive mapping, the significance of the B-mode detection was destroyed, falling from $8.22\sigma$ for the optimal and 1-s Descart algorithms to $0.02\sigma$.  This is a conclusion specific to the single detector case, as the atmospheric noise is completely correlated between detector pairs that share the same horn. Exploiting this correlation, by combining data from both channels, can completely remove the atmospheric noise.

The detector noise simulations maps also saw significant improvement, with the 1-s baseline Descart maps showing $96\%$ of the possible improvement in T and $94.71\%$ of the possible improvement in Q residuals, though the latter amounts to less than $0.1\%$ of the noise power in this case.  Whilst this is not reducible by detector time-stream differencing, the effect on the B-mode noise was very small.

The 1-s Descart maps were returned with improvements in computing time of a factor of 6 for both detector and atmospheric noise.  However, for the atmospheric noise simulations, the $98.73\%$ of the improvement in T and $99.9\%$ of Q and U was returned using a longer baseline of 10-s, which delivered an improvement in speed of $10 \times$.  Most significantly, $89.16\%$ of the improvement in Q residuals were achieved with a very long baseline of 1200-s, returning a speed improvement of $22\times$ over the maximum likelihood algorithm.  This can be understood by noting that the atmospheric noise has a lot more noise power at low frequencies than the detector noise simulations do.  As the resolution of $C_{a}$ is increased, improvements in the noise are reached more rapidly as the extra noise is resolved.

The destriping algorithm was shown not to distort the CMB signal, which is the prohibitive drawback of fast filtering methods, such as the $C_{N}^{-1/2}$ filter, which introduces extra variance into the power spectra.  Further, destriping is unbiased for B-modes, producing correct null detections for simulations with artificially zero B-modes.

The ideally cross-linked fence scan showed considerably quicker reduction in residuals for T with decreasing baseline length, than did the sabre scan with detector noise, despite identical noise in the two sets of simulations.  Improvement in residuals reached a trough at 10-s baselines: further improvements were negligible for shorter baselines.  Q and U residuals improved similarly to the sabre scan.

Covariant destriping produces very close to optimal maps, but much more quickly than the full maximum-likelihood algorithm, at speeds that make it applicable to the large datasets from upcoming B-mode experiments.  Destriping can also reduce the vast majority of atmospheric noise in single detector time-streams very quickly, using more approximate long baselines.

This paper is the first in a planned series investigating destriping for ground based polarisation experiments.  The focus of future work will be for experiments in which the Q and U signals are subject to $1/f$ noise (eg: QUIET).  For these experiments, the reduction in $1/f$ will be critically important to the science return.  The simulations will be extended to higher resolution and will investigate the effects of sub-pixel signal gradients by using CMB input maps at much higher resolution than the recovered maps.  The simulations will also be extended to multiple detectors, with noise correlated between detectors sourced both from detector fluctuations and from the atmosphere.  Destriping will be applied alongside and directly compared to the only realistic alternative: the Monte-Carlo based MASTER method where the time-stream is high-pass filtered prior to map-making \citep{hivon:2002}.

\emph{Acknowledgments}: We are grateful to Hans Kristian Eriksen, Anthony Challinor, Torsti Poutanen, Chris North, Joe Zuntz, Giancarlo de Gasperis and the C$_{\ell}$OVER collaboration for useful discussions, and to Jonathan Patterson for invaluable computing support.  We acknowledge the use of the HEALPix and FFTW subroutine libraries.  David Sutton acknowledges the support of a Science and Technology Facilities Council PhD Studentship.

\appendix

\section{Pure pseudo-$C_{l}$ estimation}
\label{pseudo Cl appendix}

Spatial information about the comparative map-making residuals is gained by analysing the angular power spectrum of the residual maps.
Estimating B-mode power spectra with partial sky coverage is a nontrivial problem, due largely to the presence of ambiguous modes \citep{bunn:2002,challinor:chon:2005}
which receive contributions from both E-mode and B-mode power on the full sky.
This ``$E\rightarrow B$ mixing'' can act as an extra source of noise, for a B-mode power spectrum estimator which does not filter out ambiguous modes.
Because B-modes are a primary scientific target for the next-generation polarisation experiments considered in this paper, it will be important for the analysis pipelines of
such experiments to use a B-mode power spectrum estimator which does not suffer from $E\rightarrow B$ mixing.
Accordingly, we analyse our residual maps using one such estimator: the pure pseudo-$C_\ell$ estimator from \citep{ksmith:2006,ksmith:2007}, which
generalizes the MASTER construction \citep{hivon:2002,brown:2005} by eliminating $E\rightarrow B$ mixing from ambiguous modes.
In this appendix, we briefly summarize its construction and key properties.

First, in order to apply pure pseudo-$C_\ell$ power spectrum estimation, one must choose (heuristically) a pixel weight function $W(\hat{n})$.
In this paper we have used cosine apodization: in spherical coordinates with north pole at the center of the survey, the weight function is given by:
\begin{equation}
W(\theta, \phi) = \left\{
\begin{array}{cl}
1 &   \theta < r-r_{*}\\
\frac{1}{2} - \frac{1}{2} \cos(\pi \frac{r - \theta}{r_{*}}) &   r-r_{*} \leq \theta \leq r\\
0 &   \theta > r
\end{array} \right.
\end{equation}
where $r$ is the survey radius and $r_{*}$ is an apodization length.  In this paper, the survey radius is $r= 7^{o}$ and heuristic apodisation radii of $r_{*} = 4.3^{o}$ and $r_{*}= 1.61^{o}$ are used for multipole ranges $\ell \le 40$ and $ \ell > 40$ respectively.

We then define pseudo multipoles $\widetilde a_{\ell m}^E, \widetilde a_{\ell m}^B$ by:
\ba
\widetilde a_{\ell m}^E &=& -\frac{1}{2} \sqrt{\frac{(l-2)!}{(l+2)!}} \int d^2\hat{n} \bigg[ \Pi_+(\hat{n}) W(\hat{n})\, \bar{\eth}\bar{\eth} Y_{\ell m}^*(\hat{n}) \nonumber \\
&& \hskip 0.9in + \Pi_-(\hat{n}) W(\hat{n}) \eth\eth Y_{\ell m}^*(x) \bigg]  \label{eq:almedef} \\
\widetilde a_{\ell m}^B &=& -\frac{i}{2} \sqrt{\frac{(l-2)!}{(l+2)!}} \int d^2\hat{n} \bigg[ \Pi_+(\hat{n}) \bar{\eth}\bar{\eth} \big( W(\hat{n}) Y_{\ell m}^*(\hat{n}) \big) \nonumber \\
&& \hskip 0.9in - \Pi_-(\hat{n}) \eth\eth \big( W(\hat{n}) Y_{\ell m}^*(x) \big) \bigg] \label{eq:almbdef}
\ea
where $\Pi_\pm(\hat{n}) = (Q\pm iU)(\hat{n})$ and $\eth, \bar{\eth}$ are the spin raising and lowering operators defined in \cite{zaldarriaga:seljak:1997}.
The derivative operators have been placed differenly in the definitions of $\widetilde a_{\ell m}^E, \widetilde a_{\ell m}^B$ above,
so that $\widetilde a_{\ell m}^E$ will receive contributions from ambiguous modes, but $\widetilde a_{\ell m}^B$ will not.
Note that $\Pi_\pm$ is a spin $(\pm 2)$ field, but the integrands appearing in Eqs.~(\ref{eq:almedef}),~(\ref{eq:almbdef})
are spin-zero and do not depend on the choice of local frame.

The next step is to define pseudo power spectra, in bands $b$, by:
\ba
\widetilde C_b^{EE} &=& \frac{1}{2\ell+1} \sum_{\ell m} P_{b\ell} |\widetilde a_{\ell m}^E|^2  \\
\widetilde C_b^{BB} &=& \frac{1}{2\ell+1} \sum_{\ell m} P_{b\ell} |\widetilde a_{\ell m}^B|^2
\ea
where $P_{b\ell}$ is a binning operator ($P_{b\ell}=0$ unless $\ell\in b$).

As constructed, the pseudo spectra are biased estimators of the signal power spectra; one has
\begin{equation}
\left( \begin{array}{c}
\langle \widetilde C_b^{EE} \rangle \\
\langle \widetilde C_b^{BB} \rangle
\end{array} \right)
=
\left( \begin{array}{cc}
K_{bb'}^+ & K_{bb'}^- \\
   0  & K_{bb'}^{+pure}
\end{array} \right)
\left( \begin{array}{c}
  C_{b'}^{EE}  \\
  C_{b'}^{BB} 
\end{array} \right)  \label{eq:transfer_matrix}
\end{equation}
where the transfer matrices $K_{bb'}$ depend only on the survey geometry and pixel weighting,
and can be computed efficiently using the algorithm in \citep{ksmith:2006}.
Note that the lower left block in Eq.~(\ref{eq:transfer_matrix}) is zero because $\widetilde a^B_{\ell m}$ does not
receive contributions from ambiguous modes.
We do not include an additive term from noise bias on the right-hand side of Eq.~(\ref{eq:transfer_matrix}) as this equation is for the case of estimating the power spectrum of a pure noise map.

The final step in the construction is to define unbiased estimators $\widehat C_b^{EE}, \widehat C_b^{BB}$ by:
\begin{equation}
\left( \begin{array}{c}
  \widehat C_b^{EE} \\
  \widehat C_b^{BB} 
\end{array} \right)
=
\left( \begin{array}{cc}
K_{bb'}^+ & K_{bb'}^- \\
   0  & K_{bb'}^{+pure}
\end{array} \right)^{-1}
\left( \begin{array}{c}
  \widetilde C_{b'}^{EE} \\
  \widetilde C_{b'}^{BB}
\end{array} \right)
\end{equation}
These are unbiased ($\langle \widehat C_b^{EE} \rangle = C_b^{EE}$ and $\langle \widehat C_b^{BB} \rangle = C_b^{BB}$) power
spectrum estimators, with no $E\rightarrow B$ mixing: the B-mode estimator has the property that it receives no contributions
from ambiguous modes.

\section{Destriping for multiple detectors including correlations}
\label{multi-detector appendix}

The time ordered output from each detector , $y_{t}$, is stacked end to end to form a single $N_{tod} \times N_{detector}$ vector.  The vector of offset function amplitudes $a_{\alpha}$, where $\alpha$ indexes the offset function, is similarly stacked end to end to form a $N_{offsets} \times N_{detector}$ vector, as are the projection matrices $P_{tp}$ and $F_{t\alpha}$.

Each time ordered datum is now indexed by $i= lt$, where $l$ denotes a particular detector and $t$ denotes a particular time.  Likewise, we index the offsets as $j= l\alpha$.  With the new indices, the full focal plane TOD vector $y_{i}$ is modelled as

\begin{equation}
y_{i}= P_{ip} x_{p} + F_{ij} a_{j} + n_{W,i}
\end{equation}
where $F_{ij} a_{j}$ models the correlated noise component of $y_{i}$ and $n_{W,i}$ is the uncorrelated white noise.

The likelihood for the system to be minimised is

\begin{eqnarray}
-2 \ln L & = & (y_{i} - P_{ip} x_{p} - F_{ij} a_{j})^{T} C_{W, ii'}^{-1} \nonumber\\
&&(y_{i'} - P_{i'p} x_{p} - F_{i'j'} a_{j'}) + a_{j}^{T} C_{a, jj'}^{-1} a_{j'}
\end{eqnarray}

The white noise covariance matrix $C_{W,ii'}$ is $n_{TOD}$ independent $N_{detector}^{2}$ matrices describing the correlation of the white noise between detectors.  For atmospheric common mode noise, we assume that this matrix is diagonal

\begin{equation}
C_{W, ii'}= \sigma^{2} \delta_{ii'}
\end{equation}
so that white noise is uncorrelated between detectors.

The offset covariance matrix $C_{a}$ encodes prior information on the correlations in the long term atmospheric $1/f$ between detectors.  Each $(N_{offset}\times N_{offset})$ sub-matrix is not diagonal but circulant and can be inverted in the Fourier domain, using the method of \cite{patanchon:2007}.

We define a multi-detector Fourier transform operator $F$, such that the Fourier transforms $\tilde{y_{k}}$ of the TOD from each detector, stacked end to end, is

\begin{equation}
\tilde{y_{k}} = F y_{i}
\label{multi FFt definition}
\end{equation}
where the index $k= l f$ denotes the Fourier mode $f$ of detector $l$.  $F$ is a block diagonal matrix, in which each block is Fourier transform operator for a single detector channel.

The Fourier domain offset covariance matrix is

\begin{equation}
R_{kk'}= F C_{a, jj'} F^{\dagger}
\label{Ca to fourier}.
\end{equation}

Each detector-detector sub-matrix of $R$, which we label $R_{ll'}$ corresponding to detector combination $ll'$, is diagonal and each diagonal element is an independent Fourier mode.  Its inverse $R^{-1}$ is easy and quick to compute explicitly, an operation that only needs to be accomplished once.

The inverse of the offset covariance is obtained by

\begin{equation}
[C_{a}^{-1}]_{ll'} = F^{\dagger} [R^{-1}]_{ll'} F
\label{fourier to Camin1}.
\end{equation}

However, there is no need to calculate and store this in the time domain, as the operation $C_{a}^{-1} a$ is completed more quickly by switching between time and Fourier space

\begin{equation}
C_{a}^{-1} a = \mathcal{F}^{-1}[ R^{-1} \mathcal{F}[a] ]
\label{multi detector Ca inversion}.
\end{equation}

Armed with this technique for painlessly inverting the offset covariance matrix, we can build the estimator for the offset amplitudes

\begin{equation}
(\mathbf{F_{j'i'} Z_{i'i} F_{ij} + \sigma_{w}^{2} C_{a, j'j}^{-1}})a_{j} = \mathbf{F_{j'i'} Z_{i'i} }y_{i}
\end{equation}
where $Z_{ii'}$ is block-diagonal in $l$ and each detector's block diagonal sub-matrix is given by (\ref{definition of Z}).

\label{lastpage}


\begin{thebibliography}{}
\bibitem[\protect\citeauthoryear{Amblard \& Hamilton}%
{2004}]{amblard:hamilton:2004} Amblard A., Hamilton J-Ch., 2004, A\&A, 417, 1189
\bibitem[\protect\citeauthoryear{Ashdown et al.}%
{2007a}]{ashdown:2007a} Ashdown M. A. J. et al., 2007a, A\&A, 467, 761
\bibitem[\protect\citeauthoryear{Ashdown et al.}%
{2007b}]{ashdown:2007b} Ashdown M. A. J. et al., 2007b, A\&A, 471, 361
\bibitem[\protect\citeauthoryear{Barret et al.}%
{2006}]{barret:2006} Barret R. et al., 2006, Templates for the Solution of Linear Systems: 
Building Blocks for Iterative Methods, 2nd Edition, SIAM
\bibitem[\protect\citeauthoryear{Borrill}%
{1999}]{borrill:1999} Borrill J., 1999,  pre-print (astro-ph/9911389v1)
\bibitem[\protect\citeauthoryear{Brown, Castro \& Taylor}%
{2005}]{brown:2005} Brown M. L., Castro P. G., Taylor A. N., 2005, MNRAS, 360, 1262
\bibitem[\protect\citeauthoryear{Bunn}%
{2002}]{bunn:2002} Bunn E. F., 2002, Phys. Rev. D, 65, 043003
\bibitem[\protect\citeauthoryear{Burigana et al.}%
{1997}]{burigana:1997} Burigana C., Malaspina M., Mandolesi N., Danese L., Maino D., Bersanelli M., Maltoni M.,
 1997, pre-print (astro-ph/9906360)
\bibitem[\protect\citeauthoryear{Bussmann, Holzapfel \& Kuo}%
{2005}]{bussman:2005}  Bussmann R. S., Holzapfel W. L., Kuo C. L., ApJ, 622,1343
\bibitem[\protect\citeauthoryear{Challinor \& Chon}%
{2005}]{challinor:chon:2005} Challinor A., Chon G., 2005, MNRAS, 360, 509
\bibitem[\protect\citeauthoryear{de Gasperis et al.}%
{2005}]{de-gasperis:2005} de Gasperis G., Balbi A., Cabella P., Natoli P., Vittorio N., 2005, A\&A, 436, 1159
\bibitem[\protect\citeauthoryear{Delabrouille}%
{1998}]{delabrouille:1998} Delabrouille J., 1998, A\&AS, 127, 555
\bibitem[\protect\citeauthoryear{Dore et al.}%
{2001}]{dore:2001} Dore O., Teyssier R., Bouchet F.R., Vibert D., Prunet S., 2001, A\&A, 374, 358
\bibitem[\protect\citeauthoryear{Ferreira \& Jaffe}%
{2000}]{ferreira:jaffe:2000} Ferreira P. G., \& Jaffe A. H., 2000, MNRAS, 312, 89
\bibitem[\protect\citeauthoryear{Gorski et al.}%
{2005}]{gorski:2005} Gorski K. M., Hivon E., Banday A. J., Wandelt B. D., Hansen F. K., Reinecke M., Bartelmann M., 2005, ApJ, 622, 759
\bibitem[\protect\citeauthoryear{Hanany \& Rosenkranz}%
{2003}]{hanany:2003} Hanany S., Rosenkranz P., 2003, NewAR, 47, 1159
\bibitem[\protect\citeauthoryear{Hanany et al.}%
{2000}]{hanany:2000} Hanany S. et al. 2000, ApJ, 545, L5
\bibitem[\protect\citeauthoryear{Hinderks et al.}%
{2008}]{hinderks:2008} Hinderks J. et al., 2008, pre-print (astro-ph/0805.1990v1)
\bibitem[\protect\citeauthoryear{Hinshaw et al.}%
{2008}]{hinshaw:2008} Hinshaw G. et al. 2008, pre-print (astro-ph/0803.0732)
\bibitem[\protect\citeauthoryear{Hivon et al.}%
{2002}]{hivon:2002} Hivon E., Gorski K. M., Netterfield C. B., Crill P. C., Prunet S., Hansen F., 2002, ApJ, 567, 2
\bibitem[\protect\citeauthoryear{Johnson et al.}{2007}]%
{johnson:2007} Johnson B. R. et al., 2007, ApJ, 665, 42J
\bibitem[\protect\citeauthoryear{Keihanen et al.}%
{2004}]{keihanen:2004} Keihanen E., Kurki-Suonio H., Poutanen T., Maino D., Burigana C., 2004 A\&A, 428, 287
\bibitem[\protect\citeauthoryear{Keihanen et al.}%
{2005}]{keihanen:2005} Keihanen E., Kurki-Suonio H., Poutanen T., 2005, MNRAS, 360, 390
\bibitem[\protect\citeauthoryear{Lewis, Challinor \& Lasenby}%
{2000}]{lewis:2000} Lewis A., Challinor A., Lasenby A., 2000, ApJ, 538, 473
\bibitem[\protect\citeauthoryear{Maino et al. }%
{1999}]{maino:1999}  Maino D. et al., 1999, A\&AS, 140, 383
\bibitem[\protect\citeauthoryear{MacTavish et al. }%
{2007}]{mactavish:2007}  MacTavish C. J. et al., 2007, pre-print (astro-ph/0710.0375)
\bibitem[\protect\citeauthoryear{Macias-Perez et al.}%
{2007}]{macias-perez:2007}  Macias-Perez J. F. et al., 1999, A\&A, 467, 1313
\bibitem[\protect\citeauthoryear{Masi et al. }%
{2006}]{masi:2006}  Masi S. et al., 2006, ApJ, 458, 687
\bibitem[\protect\citeauthoryear{Natoli et al.}%
{2001}]{natoli:2001} Natoli P., de Gasperis G., Gheller C., Vittorio N., 2007, A\&A, 372, 346
\bibitem[\protect\citeauthoryear{North et al.}%
{2008}]{north:2008} North C. E. et al., 2008, pre-print (astro-ph/0805.3690)
\bibitem[\protect\citeauthoryear{Oxley et al.}%
{2005}]{oxley:2005} Oxley P. et al., 2005, pre-print (astro-ph/0501111)
\bibitem[\protect\citeauthoryear{Patanchon et al.}%
{2007}]{patanchon:2007} Patanchon G. et al., 2007, pre-print (astro-ph/0711.3462)
\bibitem[\protect\citeauthoryear{Poutanen et al.}%
{2006}]{poutanen:2006} Poutanen et al., 2006, A\&A, 449,1311
\bibitem[\protect\citeauthoryear{Press et al.}%
{2002}]{press:2002} Press et al., 2002, Numerical Recipes in C++, 2nd Edition, Cambridge 
\bibitem[\protect\citeauthoryear{Smith}%
{2006}]{ksmith:2006} Smith K. M., 2006, Phys. Rev. D, 74, 083002
\bibitem[\protect\citeauthoryear{Smith \& Zaldarriaga}%
{2007}]{ksmith:2007} Smith K. M., Zaldarriaga M., 2007, Phys. Rev. D, 76, 043001
\bibitem[\protect\citeauthoryear{Smoot et al.}%
{1992}]{smoot:1992} Smoot G. F. et al., 1992, ApJ, 396, L1
\bibitem[\protect\citeauthoryear{Stompor et al.}%
{2002}]{stompor:2002} Stompor R. et al., 2002, Phys. Rev. D, 65, 022003
\bibitem[\protect\citeauthoryear{Tegmark}%
{1997a}]{tegmark:1997a} Tegmark M., 1997a, ApJ, 480, L87-L90
\bibitem[\protect\citeauthoryear{Tegmark}%
{1997b}]{tegmark:1997b} Tegmark M., 1997b, Phys. Rev. D, 56,8,4514
\bibitem[\protect\citeauthoryear{Zaldarriaga, \& Seljak}%
{1997}]{zaldarriaga:seljak:1997} Zaldarriaga M., Seljak U., 1997, Phys. Rev. D, 55, 4 1830
\end{thebibliography}
\end{document}